\RequirePackage{fix-cm} 
\documentclass[a4paper, twoside, reqno, dvips, 12pt]{amsart}
\usepackage{fixltx2e}   

\usepackage{etex}

\usepackage[ugly]{nicefrac}

\usepackage[latin1]{inputenc}
\usepackage[T1]{fontenc}

\usepackage{algorithm}
\usepackage[noend]{algpseudocode}

\usepackage[titletoc,title]{appendix}

\usepackage{esint}
\usepackage{dsfont}
\usepackage{xspace}
\usepackage{amsgen}
\usepackage{amsthm}
\usepackage{amssymb}
\usepackage{amsmath}
\usepackage{wasysym}
\usepackage{upgreek}
\usepackage{amsfonts}
\usepackage{stmaryrd}
\usepackage{mathtools}

\usepackage{relsize}
\usepackage{textcomp}
\usepackage{textgreek}
\usepackage[mathcal, mathscr]{euscript}

\usepackage{mathrsfs}
\DeclareMathAlphabet{\mathscrbf}{OMS}{mdugm}{b}{n}

\usepackage{a4wide}

\usepackage[dvipsnames, table]{xcolor}
\definecolor{bckg}{RGB}{20.8, 20.8, 20.8}
\definecolor{oneblue}{rgb}{0.0, 0.0, 0.85}
\definecolor{Lightblue}{RGB}{214, 214, 214}
\definecolor{bluepigment}{rgb}{0.2, 0.2, 0.6}
\definecolor{charcoal}{rgb}{0.21, 0.27, 0.31}
\definecolor{denimblue}{rgb}{0.08, 0.38, 0.74}
\definecolor{Lightgray}{rgb}{0.89, 0.89, 0.89}
\definecolor{darkgrey}{rgb}{0.273, 0.281, 0.30}
\definecolor{darkelectricblue}{rgb}{0.33, 0.41, 0.47}

\usepackage{multirow}

\usepackage[sort&compress, comma, square, numbers]{natbib}

\usepackage{psfrag}
\usepackage{graphicx}
\usepackage{adjustbox}
\usepackage[tight]{subfigure}
\usepackage{morefloats}
\usepackage{indentfirst}

\usepackage[usenames, dvipsnames, pdf]{pstricks}
\usepackage{epsfig}
\usepackage{pst-grad} 
\usepackage{pst-plot} 

\usepackage{rotating}
\usepackage{pdflscape}

\usepackage{acronym}
\usepackage{microtype}
\usepackage[labelsep=period,%
            labelfont={bf,sf,color=bluepigment},%
            justification=raggedright]{caption}

\usepackage[perpage, symbol]{footmisc}

\usepackage[colorlinks,
           urlcolor=oneblue,
           linkcolor=denimblue,
           citecolor=NavyBlue,
           bookmarksopen=false,
           pdfpagemode=UseNone,
           pagebackref]{hyperref}

\usepackage[explicit]{titlesec}

\titleformat{\section}[block]
  {\color{NavyBlue}\Large\sffamily\bfseries}
  {}
  {0.0em}
  {\colorbox{bckg!5}{\strut\parbox{\dimexpr\linewidth-2\fboxsep\relax}{\thesection. #1}}}
  [\vspace*{0.33em}]

\titleformat{name=\section,numberless}[block]
  {\color{NavyBlue}\Large\sffamily\bfseries}
  {}
  {0.0em}
  {\colorbox{bckg!5}{\strut\parbox{\dimexpr\linewidth-2\fboxsep\relax}{#1}}}
  [\vspace*{0.33em}]

\titleformat{\subsection}
  {\color{NavyBlue}\large\sffamily\bfseries}
  {}
  {0.0em}
  {\colorbox{bckg!5}{\parbox{\dimexpr\linewidth-2\fboxsep\relax}{\thesubsection. #1}}}
  [\vspace*{0.33em}]

\titleformat{name=\subsection,numberless}
  {\color{NavyBlue}\Large\sffamily\bfseries}
  {}
  {0em}
  {\colorbox{bckg!5}{\parbox{\dimexpr\linewidth-2\fboxsep\relax}{#1}}}
  [\vspace*{0.33em}]

\titleformat{\subsubsection}
  {\color{bluepigment}\sffamily\normalsize\bfseries}
  {\thesubsubsection}
  {0.5em}
  {#1}
  [\vspace*{0.33em}]

\titleformat{\paragraph}[runin]
  {\color{bluepigment}\sffamily\small\bfseries}
  {}
  {0em}
  {#1}

\titlespacing{\section}{0.0em}{1.5em plus 2pt minus 2pt}%
{1.0em plus 2pt minus 2pt}[0em]
\titlespacing{\subsection}{0.5em}{1.5em plus 2pt minus 2pt}%
{1.0em}[0em]
\titlespacing{\subsubsection}{0.5em}{1.5em plus 2pt minus 2pt}%
{1.0em plus 2pt minus 2pt}[0em]

\usepackage{titletoc}

\contentsmargin{0.5em}
\newlength{\tocsep} 
\titlecontents{section}[\tocsep]
  {\addvspace{10pt}\bfseries\sffamily}
  {\contentslabel[\thecontentslabel]{\tocsep}}
  {}
  {\ \titlerule*[0.75pc]{.}\ \thecontentspage}
  []
\titlecontents{subsection}[\tocsep]
  {\addvspace{8pt}\sffamily}
  {\contentslabel[\thecontentslabel]{\tocsep}}
  {}
  {\ \titlerule*[0.5pc]{.}\ \thecontentspage}
  []
\titlecontents*{subsubsection}[\tocsep]
  {\addvspace{2pt}\footnotesize\sffamily}
  {}
  {}
  {\ \titlerule*[0.35pc]{.}\ \thecontentspage}
  [\\*]

\makeatletter
\def\@setauthors{%
  \begingroup
  \def\thanks{\protect\thanks@warning}%
  \trivlist
  \centering\footnotesize \@topsep30\p@\relax
  \advance\@topsep by -\baselineskip
  \item\relax
  \author@andify\authors
  \def\\{\protect\linebreak}%
  \textsc{\normalsize\textcolor{darkelectricblue}{\authors}}%
  \ifx\@empty\contribs
  \else
    ,\penalty-3 \space \@setcontribs
    \@closetoccontribs
  \fi
  \endtrivlist
  \endgroup
}
\def\@settitle{\begin{center}%
  \baselineskip14\p@\relax
    \bfseries
    \textsc{\Large\textcolor{charcoal}{\@title}}
  \end{center}%
}
\makeatother

\usepackage{enumitem}
\setlist[description]{%
  topsep=30pt,               
  itemsep=5pt,               
  font={\bfseries\sffamily\color{NavyBlue}}, 
}

\usepackage{fancyhdr}
\usepackage{lastpage}

\newcommand*\Title{\textcolor{bluepigment}{An efficient numerical model for liquid water uptake}}
\newcommand*\Authors{\textcolor{bluepigment}{A.~Jumabekova, J.~Berger, D.~Dutykh \etal}}
\newcommand*{\plogo}{\textcolor{gray}{{\texttt{arXiv.org} / \textsc{hal}}}} 

\numberwithin{equation}{section}

\newcommand{\etal}{\emph{et al.}\xspace}

\newcommand{\ue}{\mathrm{e}}
\newcommand{\ud}{\text{d}}
\newcommand*\vit{\mathsf{v}}
\renewcommand{\leq}{\leqslant}
\DeclareMathOperator{\scd}{scd} 
\let\norm\relax
\DeclarePairedDelimiterX{\norm}[1]{\lVert}{\rVert}{#1}

\newcommand{\scal}{\boldsymbol{\cdot}}
\newcommand*\od[2]{\frac{\mathrm{d} #1}{\mathrm{d} #2}}
\newcommand*\pd[2]{\frac{\partial #1}{\partial #2}}
\renewcommand{\div}{\grad\scal}
\newcommand{\grad}{\boldsymbol{\nabla}}
\newcommand{\eqdef}{\mathop{\stackrel{\,\mathrm{def}}{:=}\,}}

\renewcommand{\O}{\mathcal{O}}

\newcommand{\Eu}{\textsc{Euler}}
\newcommand{\SG}{\textsc{Scharfetter}--\textsc{Gummel}}
\newcommand{\Fo}{\mathrm{Fo}}
\newcommand{\Pe}{\mathrm{Pe}}
\newcommand{\Bo}{\mathrm{Bo}}
\newcommand{\uinf}{u_{\,\mathrm{inf}}}

\newcommand{\half}{{\textstyle{1\over2}}}

\newcommand*\egal{\ = \ }
\newcommand*\plus{\ + \ }
\newcommand*\moins{\ - \ }

\begin{document}

\title[\Title]{An efficient numerical model for liquid water uptake in porous material and its parameter estimation}

\author[A.~Jumabekova]{Ainagul Jumabekova}
\address{\textbf{A.~Jumabekova:} Univ. Grenoble Alpes, Univ. Savoie Mont Blanc, UMR 5271 CNRS, LOCIE, 73000 Chamb\'ery, France}
\email{Ainagul.Jumabekova@univ-smb.fr}
\urladdr{https://www.researchgate.net/profile/Ainagul\_Jumabekova/}

\author[J.~Berger]{Julien Berger}
\address{\textbf{J.~Berger:} Univ. Grenoble Alpes, Univ. Savoie Mont Blanc, UMR 5271 CNRS, LOCIE, 73000 Chamb\'ery, France}
\email{Berger.Julien@univ-smb.fr}
\urladdr{https://www.researchgate.net/profile/Julien\_Berger3/}

\author[D.~Dutykh]{Denys Dutykh}
\address{\textbf{D.~Dutykh:} Univ. Grenoble Alpes, Univ. Savoie Mont Blanc, CNRS, LAMA, 73000 Chamb\'ery, France and LAMA, UMR 5127 CNRS, Universit\'e Savoie Mont Blanc, Campus Scientifique, F-73376 Le Bourget-du-Lac Cedex, France}
\email{Denys.Dutykh@univ-smb.fr}
\urladdr{http://www.denys-dutykh.com/}

\author[H.~Le~Meur]{Herv\'e Le Meur}
\address{\textbf{H.~Le~Meur:} Laboratoire de Math\'ematiques d'Orsay, Univ. Paris-Sud, CNRS, Universit\'e Paris-Saclay, 91405 Orsay, France}
\email{Herve.LeMeur@math.u-psud.fr}
\urladdr{https://www.math.u-psud.fr/\char`~lemeur/}

\author[A.~Foucquier]{Aur\'elie Foucquier}
\address{\textbf{A.~Foucquier:} Univ. Grenoble Alpes, CEA, LITEN, DTS, INES, F-38000, Grenoble, France}
\email{Aurelie.Foucquier@gmail.com}
\urladdr{https://www.researchgate.net/profile/Aurelie\_Foucquier/}

\author[M.~Pailha]{Mickael Pailha}
\address{\textbf{M.~Pailha:} Univ. Grenoble Alpes, Univ. Savoie Mont Blanc, UMR 5271 CNRS, LOCIE, 73000 Chamb\'ery, France}
\email{Mickael.Pailha@univ-smb.fr}

\author[Ch.~M\'en\'ezo]{Christophe M\'en\'ezo}
\address{\textbf{Ch.~M\'en\'ezo:} Univ. Grenoble Alpes, Univ. Savoie Mont Blanc, UMR 5271 CNRS, LOCIE, 73000 Chamb\'ery, France}
\email{Christophe.Menezo@univ-smb.fr}
\urladdr{https://www.researchgate.net/profile/Christophe\_Menezo/}


\begin{titlepage}
\thispagestyle{empty} 
\noindent
{\Large Ainagul \textsc{Jumabekova}}\\
{\it\textcolor{gray}{LOCIE--CNRS, Universit\'e Savoie Mont Blanc, France}}
\\[0.02\textheight]
{\Large Julien \textsc{Berger}}\\
{\it\textcolor{gray}{LOCIE--CNRS, Universit\'e Savoie Mont Blanc, France}}
\\[0.02\textheight]
{\Large Denys \textsc{Dutykh}}\\
{\it\textcolor{gray}{LAMA--CNRS, Universit\'e Savoie Mont Blanc, France}}
\\[0.02\textheight]
{\Large Herv\'e \textsc{Le Meur}}\\
{\it\textcolor{gray}{Universit\'e Paris-Saclay, Orsay, France}}
\\[0.02\textheight]
{\Large Aur\'elie \textsc{Foucquier}}\\
{\it\textcolor{gray}{CEA--LITEN, DTS, France}}
\\[0.02\textheight]
{\Large Mickael \textsc{Pailha}}\\
{\it\textcolor{gray}{LOCIE--CNRS, Universit\'e Savoie Mont Blanc, France}}
\\[0.02\textheight]
{\Large Christophe \textsc{M\'en\'ezo}}\\
{\it\textcolor{gray}{LOCIE--CNRS, Universit\'e Savoie Mont Blanc, France}}
\\[0.04\textheight]

\colorbox{Lightblue}{
  \parbox[t]{1.0\textwidth}{
    \centering\huge\sc
    \vspace*{0.7cm}
    
    \textcolor{bluepigment}{An efficient numerical model for liquid water uptake in porous material and its parameter estimation}

    \vspace*{0.7cm}
  }
}

\vfill 

\raggedleft     
{\large \plogo} 
\end{titlepage}


\newpage
\thispagestyle{empty} 
\par\vspace*{\fill}   
\begin{flushright} 
{\textcolor{denimblue}{\textsc{Last modified:}} \today}
\end{flushright}


\newpage
\tableofcontents
\thispagestyle{empty}


\begin{abstract}

The goal of this study is to propose an efficient numerical model for the predictions of capillary adsorption phenomena in a porous material. The \textsc{Scharfetter--Gummel} numerical scheme is proposed to solve an advection-diffusion equation with gravity flux. Its advantages such as accuracy, relaxed stability condition, and reduced computational cost are discussed along with the study of linear and nonlinear cases. The reliability of the numerical model is evaluated by comparing the numerical predictions with experimental observations of liquid uptake in bricks. A parameter estimation problem is solved to adjust the uncertain coefficients of moisture diffusivity and hydraulic conductivity.


\bigskip\bigskip
\noindent \textbf{\keywordsname:} water uptake process; porous material; \SG ~numerical scheme; parameter estimation problem; advection--diffusion equation with gravity flux \\


\end{abstract}


\newpage
\section{Introduction}

Nowadays efficient energy consumption of buildings is one of the most challenging tasks. In case of the historical buildings, this problem has to deal not only with the energy performance but also with the building protection itself \cite{Torres2007, Falchi2018}. The study of moisture's amount in the building walls takes the largest part for energy reduction since it can raise the heat loss through the walls. The increase of moisture quantity may occur due to different reasons as wind--driven rain, rising damp or simply vapour diffusion \cite{Berger2015a}.

Among all the potential sources of moisture, rising damp is a major issue since it is an important source of liquid water. It may strongly impact the building structure by modifying the mechanical behaviour of the wall. Moreover, the water from the ground brings dissolved salts. During the evaporation process, these salts precipitate and crystals appear that deteriorate the material \cite{Scherer1999, Scherer2004}. Some examples of such deterioration can be found in \cite{Lopez-Arce2009} or \cite{Rirsch2010}.

Thus, one understands the importance of studying the physical phenomena occurring in rising damp. As mentioned by \textsc{Franzoni} in \cite{Franzoni2014}, the main challenges related to the rising damp problem is to find accurate, fast and cheap methods to characterize the moisture content during the water uptake of a material or a wall. Several methods to characterize the amount of moisture already exist and can be divided in two main groups: \textit{(i)} invasive (traditional like weighting--drying, chromatography) and \textit{(ii)}  non-invasive (dielectric, microwave) \cite{Hoa2017}. But usually, these approaches are very complex procedures requiring extensive technical support. Moreover, the obtained results do not describe an exact situation of the whole building. Thus, as underlined in \cite{Falchi2018}, it is worth investigating other approaches by proposing numerical models to perform simulations and obtain quantitative results.

Several models were proposed in the literature for the simulation of liquid water uptake in a porous material. A first approach is to propose analytical solutions of the problem as for instance in \cite{Guimaraes2010, Janetti2017}. However, the main drawback is that these solutions work only for idealized conditions. As an example, in \cite{Janetti2017} a step--function is used for the diffusivity coefficient, neglecting its nonlinear variation with the fields. This assumption is not valid for the practical case study. Other approaches are based on standard numerical methods in \cite{Haupl1997, Guizzardi2016, Peszynksa2008, Torres2010}. However, these approaches also have an important drawback. When using explicit \textsc{Euler} approach, the standard stability conditions must be respected implying very small values of discretization time grids due to the high nonlinearity of the problem. Thus these numerical models have a high computational cost. These works promote the use of implicit schemes for their unconditional stability properties. However, they require a large number of sub--iterations during the computation to treat nonlinearities, which also lead to an increase in the computational cost of the numerical models. It is of major importance to propose efficient numerical models to represent the physical phenomena of capillary adsorption in a porous material.

When proposing efficient numerical models, one important point is its reliability to represent the physical phenomena. For this, the numerical predictions need to be confronted to experimental observations. In the case of rising damp, there are several material properties involved in the definition of the physical model. To have an accurate numerical prediction, the moisture diffusivity and the hydraulic conductivity must be known precisely. As mentioned in \cite{Guizzardi2016}, there is a lack of data in the literature for these material properties. Thus, when comparing the numerical predictions to experimental observation, the material properties may be estimated to calibrate the numerical model. This procedure requires solving the parameter estimation problem. The solution of such type of the inverse problem requires the computation of the so-called direct problem several times. Thus, since current numerical models have high computational cost, the estimation of material properties becomes an obstacle.

Therefore, the objective of this article is twofold. First, an innovative numerical model is proposed for capillary adsorption phenomena in a porous material. It is based on the \SG ~numerical scheme. This approach is particularly efficient for so-called advection--diffusion equations as highlighted from a mathematical point of view in \cite{Gosse2017} and illustrated in \cite{Berger2017a, Berger2018a} for the case of heat and moisture transfer in building porous materials. In our work, the proposed numerical model is compared to the standard methods in the context of capillary adsorption phenomena. The issue is to prove its reliability, accuracy and smaller computational cost. Then, the numerical model is compared to experimental data to evaluate its accuracy to represent the physical phenomena. To calibrate the model, a parameter estimation problem is solved in a good time to determine the material properties of a brick using observed data of water uptake.

The paper is organized in the following way. In Section~\ref{seq:phys_model}, the physical model is described. Following Section~\ref{sec:Numerical_method} presents the \SG ~scheme. Its accuracy and advantages are compared to classical methods on numerical case studies. Section~\ref{sec:exp_facility} describes the experimental facility used to generate observation of water uptake. Section~\ref{sec:parameter_estimation} aims at comparing the numerical predictions and the experimental observations, as well as estimating the material properties of the brick to calibrate the model.


\section{Physical model}
\label{seq:phys_model}

\subsection{Liquid transfer}
\label{sec:liquid_transfer}

Let us present the physical model of liquid water uptake process in a porous material. The physical model is inspired by the one proposed by \textsc{Philipp} and \textsc{De Vries} in 1957 \cite{Philip1957}. Since the problem deals with high moisture content, the exchange between vapor and liquid water is supposed negligible. Moreover, it is assumed that no chemical reaction occurs. Thus, the liquid water conservation equation is given by:
\begin{align*}
  \pd{\theta}{t} \egal - \, \div \, \boldsymbol{\mathbf{j}} \,,
\end{align*}
where $\theta  \ \mathsf{\bigl[\,m^{\,3}/m^{\,3}\,\bigr]}$ is the volume basis of liquid water content in the material and \\ $\Bigl|\Bigl|\, \overrightarrow{\boldsymbol{\mathbf{j}}} \,\Bigr|\Bigr| \ \mathsf{\Bigl[\,kg/(m^{\,2}.s)\,\Bigr]}$ is the liquid flow through the capillaries.

The liquid flow is driven by diffusion, advection due to air pressure difference and gravity forces \cite{Hall2002}:
\begin{align*}
  \boldsymbol{j} \egal \boldsymbol{j}_{\,\mathrm{d}}  \plus \boldsymbol{j}_{\,\mathrm{a}} \plus \boldsymbol{j}_{\,\mathrm{g}} \,.
\end{align*}

\begin{figure}
	\centering
	\includegraphics[width=0.4\textwidth]{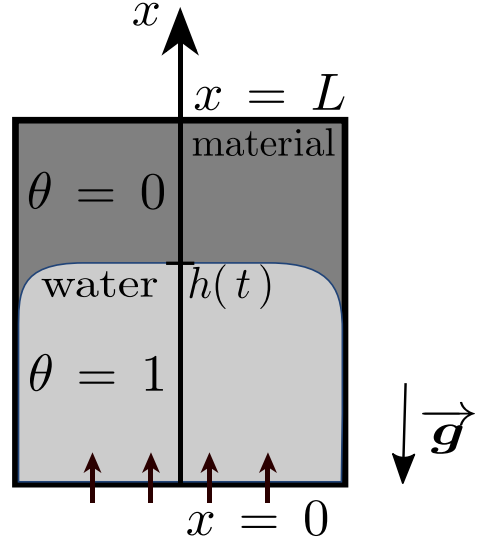}
	\caption{\small\em Illustration of the problem of water uptake in the brick.}
	\label{fig:illustration_problem}
\end{figure}

The diffusion liquid transfer flux is expressed as: 
\begin{align*}
  \boldsymbol{j}_{\,\mathrm{d}} \egal D_{\,\theta} \, \grad \theta \plus D_{\,T} \, \grad T \,,
\end{align*}
where $D_{\,\theta} \ \mathsf{\bigl[\,m^{\,2}/s\,\bigr]}$ is the liquid transport coefficient under a water content gradient and $D_{\,T} \ \mathsf{\bigl[\,m^{\,2}/(s.K)\,\bigr]}$ is the liquid transport coefficient associated to a temperature gradient. Both depend on water content $\theta\,$. Since the experiments are conducted under isothermal conditions, the liquid diffusion under temperature gradient is assumed negligible. Thus, we have:
\begin{align*}
  \boldsymbol{j}_{\,\mathrm{d}} \egal D_{\,\theta} \, \grad \theta \,.
\end{align*}
The transfer is also driven by gravity forces described by the following expression:
\begin{align*}
  \boldsymbol{j}_{\,\mathrm{g}} \egal K \, \boldsymbol{i} \,,
\end{align*}
where $K \ \mathsf{\bigl[\,m/s\,\bigr]}$ is the hydraulic conductivity depending on $\theta\,$, and $\boldsymbol{i}$ is taken as pointing upward as illustrated in Figure~\ref{fig:illustration_problem}. To take into account the motion due to the gradient pressure, an advection flux is introduced. It represents liquid water transfer induced by the filtration of dry air:
\begin{align*}
  \boldsymbol{j}_{\,\mathrm{a}} \egal \Pi \, \theta \, \boldsymbol{\vit} \,,
\end{align*}
where $\boldsymbol{\vit} \ \mathsf{\bigl[\,m/s\,\bigr]}$ is the mass average velocity and $\Pi$ the porosity of the material. Since the filtration of dry air only influences the water front at $x \egal H \,$, the mass average velocity is expressed as: 
\begin{align*}
  \boldsymbol{\vit} \egal \,\biggl(\,1\moins \frac{H}{L}\,\biggr) \, \boldsymbol{\vit}_{\,0} \,, 
\end{align*}
where $H$ is the water height in the brick, calculated as the total moisture content at the current instant $t\,$:
\begin{align*}
  H\,(\,t\,) \egal \int_{\,0}^{\,L} \, \frac{\theta\,(\,x,t\,)\,}{\theta_{\,\mathrm{sat}}}\;\mathrm{d}x \,,
\end{align*}
where  $\theta_{\,\mathrm{sat}}$ -- saturation moisture content.

Let us define coefficient $a_{\,0}$ and approximate it using \textsc{Darcy}'s law as:
\begin{align*}
  a_{\,0} \egal \Pi \, \vit_{\,0} \, \approx \, \Pi \, \frac{k_{\,\mathrm{a}}}{\mu} \, \frac{\Delta P}{L} \,,
\end{align*}
where $k_{\,\mathrm{a}} \, \mathsf{\bigl[\,m^{\,2}\,\bigr]}$ is the intrinsic permeability of the material, and $\mu \, \mathsf{\bigl[\,Pa\cdot s\,\bigr]}$ is the dynamic viscosity of the fluid.

Finally, the $1-$dimensional liquid transfer equation is expressed as: 
\begin{align}\label{eq:liquid_water_uptake_eq}
  \pd{\theta}{t} \egal \pd{}{x} \, \biggl(\, D_{\,\theta} \, \pd{\theta}{x} \moins a_{\,0}\, (\,1\moins H\,) \,\theta \, \boldsymbol{i} \moins K \, \boldsymbol{i}  \,\biggr) \,.
\end{align}

In the physical model, \textsc{Dirichlet} boundary conditions are set since the surface convective transfer coefficients are unknown. At the boundary where liquid uptakes, the water content is set to saturation. At the top of the brick, the water content remains equal to the initial condition where no liquid is present in the brick. Thus, the boundary and initial conditions are defined as:
\begin{align*}
  & \theta\,(\,x\egal0\,,t\,) \egal \begin{cases} 0 \,, &t\egal 0 \\\theta_{\,\mathrm{sat}}\,, &t\ > \ 0  \end{cases}\,, 
  && \theta\,(\,x\egal L\,,t\,) \egal 0 \,,
  && \theta\,(\,x\,,\,t\egal 0) \egal 0 \,.
\end{align*}

\subsection{Dimensionless formulation}

For numerical analysis, it is very useful to work with the unitless formulation of the model. The first reason is the computational accuracy. Scaling the problem to the appropriate order, many terms will have the same order, so the effects of numerical errors are minimized when computing the residual. The second reason is that many useful relationships exist between dimensionless numbers, which show how specific things influence the whole system.

The following dimensionless variables for the moisture content, the time and space domains are defined:
\begin{align*}
  & u \egal \frac{\theta}{\theta_{\,\mathrm{sat}}} \,, 
  && x^{\,\star} \egal \frac{x}{L} \,,
  && t^{\,\star} \egal \frac{t}{t_{\,\mathrm{ref}}} \,,
\end{align*}
where $\theta_{\,\mathrm{sat}}$ -- saturation moisture content, $L$ -- length of the brick, $t_{\,\mathrm{ref}}$ --  set to one hour.

Then all the material properties are modified considering a reference value:
\begin{align*}
  & d^{\,\star}\,(\,u\,) \egal \nicefrac{D_{\,\theta}(\,\theta\,)}{D_{\,\mathrm{ref}}}\,,
  && a_{\,0}^{\,\star}\egal\frac{a_{\,0}}{a_{\,\mathrm{ref}}}\,,
  &&& k^{\,\star}\,(\,u\,) \egal \nicefrac{K(\,\theta\,)}{k_{\,\mathrm{ref}}}\,.
\end{align*}

The \textsc{Fourier} number measures the relative importance of the heat and mass transfers through the material:
\begin{align*}
  \mathrm{Fo} \egal \nicefrac{t_{\,\mathrm{ref}}\,D_{\,\mathrm{ref}}}{L^{\,2}} \,.
\end{align*}
The \textsc{P\'eclet} number demonstrates the relative importance of the advection against the diffusion transfer:
\begin{align*}
  \mathrm{Pe} \egal \nicefrac{a_{\,\mathrm{ref}}\,L}{D_{\,\mathrm{ref}}} \,.
\end{align*}
The importance of the gravity forces with the respect to the diffusion is quantified by the \textsc{Bond} number:
\begin{align*}
  \mathrm{Bo} \egal \nicefrac{k_{\,\mathrm{ref}}\,L}{\theta_{\,\mathrm{sat}}\,D_{\,\mathrm{ref}}} \,.
\end{align*}
Finally, the following dimensionless equation of liquid transfer is obtained:
\begin{align}\label{dim_eq1}
  \pd{u}{t^{\,\star}} & \egal \mathrm{Fo}\,\pd{}{x^{\,\star}} \biggl(\, d^{\,\star}\,(\,u\,) \, \pd{u}{x^{\,\star}} \moins \mathrm{Pe} \, a_{\,0}^{\,\star} \, \Bigl(\,1 \moins H^{\,\star} \,\Bigr) \, u \moins \mathrm{Bo} \, k^{\,\star}\,(\,u\,) \,\biggr)  \,,
\end{align} 
where the dimensionless water front height is given by:
\begin{align}\label{eq:definition_H}
  H^{\,\star} \egal \int_{\,0}^{\,1} \, u\,(\,x^{\,\star}\,,t^{\,\star}\,)\, \mathrm{d}\,x^{\,\star} \,. 
\end{align}
Further in article, for the sake of clarity the upper script $^{\,\star}$ is omitted, and Eq.~\eqref{dim_eq1} transforms into:
\begin{align} \label{eq:water_uptake_dimless}
  \pd{u}{t} & \egal \Fo \, \cdot \, \pd{}{x}\, \biggl(\,d\,(\,u\,) \, \cdot \, \pd{u}{x} \moins \Pe \, \cdot \,  a\,(\,u\,)\, \cdot \, u \moins \Bo \, \cdot \,k\,(\,u\,) \,\biggr)  \,,
\end{align} 
with the following boundary and initial conditions: 
\begin{align*}
  & u(\,x\egal0\,,\,t) \egal \begin{cases} 0 \,, &t\egal 0 \\1\,, &t\ > \ 0  \end{cases}\,, 
  && u(\,x\egal1\,,\,t) \egal 0 \,,
  && u(\,x\,,\,t\egal 0) \egal 0 \,.
\end{align*}
Next section will introduce how to solve the retrieved equation effectively.


\section{Numerical method}
\label{sec:Numerical_method}

The issue is now to propose an efficient numerical model to compute an accurate solution with a reduced computational cost. For this, the \SG ~numerical scheme for an advection--diffusion equation with gravity flux is introduced. Its mathematical properties are discussed. For the sake of simplicity, the following differential equation is considered:
\begin{align}\label{eq:main_eq}
  \pd {u}{t} \ &= \ \pd{}{x}\, \Biggl(\, d\, \cdot\,\pd{u}{x} \moins a\, \cdot\,u \moins k\,(u) \,\Biggr)\,, &&  \, t \ > \ 0 \,,\quad x \ \in \ \bigl[\,0 \,,\, 1\,\bigr]\,, 
\end{align}
where $\,a\,$, $\,d\,$ are constants and coefficient $\,k\,(u)\,$ depends on $u\,$. \textsc{Dirichlet} boundary conditions are taken as 
\begin{align*}
  & u\,(\,0,t\,) \egal u_{\,L}(\,t\,) \,,
  &&  u\,(\,1,t\,) \egal u_{\,R}(\,t\,)\,.
\end{align*}

A brief demonstration of the uniqueness of the solution of Eq.~\eqref{eq:main_eq} is provided for the linearized equation in Appendix~\ref{sec:uniqueness_solution}. It can be remarked that the uniqueness of advection--diffusion equation has been demonstrated in \cite{Chen1988} considering a nonlinear advective term and a linear diffusive one with \textsc{Dirichlet}--type boundary conditions. Some demonstrations have been made for multi-dimensional systems of advection--diffusion equations also with \textsc{Dirichlet}--type boundary conditions. In \cite{Evans2010}, the problem considered is linear. In \cite{Kusano1965}, it is quasi--linear since the diffusion coefficients depend on space and time whereas the advective and source terms are nonlinear. Some proofs have been also produced for a system of diffusion equations with diffusion coefficients depending on the fields and \textsc{Neumann}--type boundary conditions in \cite{Araujo2017}.

Let us discretize uniformly the space and time intervals, with the parameters $\Delta x$ and $\Delta t\,$, respectively. The discrete values of function $u\,(x,\,t)$ are defined as $\,u_j^n \ \equiv \ u\,(\,x_j,\,t_n\,)\,$, where $\,j \, \in \, \{\,1,\,\ldots\,,N\, \}\, $ and $\,n\, \in \, \{\,1,\,\ldots\,,N_t\, \}\,$.


\subsection{The \SG ~numerical scheme}

\begin{figure}
	\centering
	\includegraphics[width=0.7\textwidth]{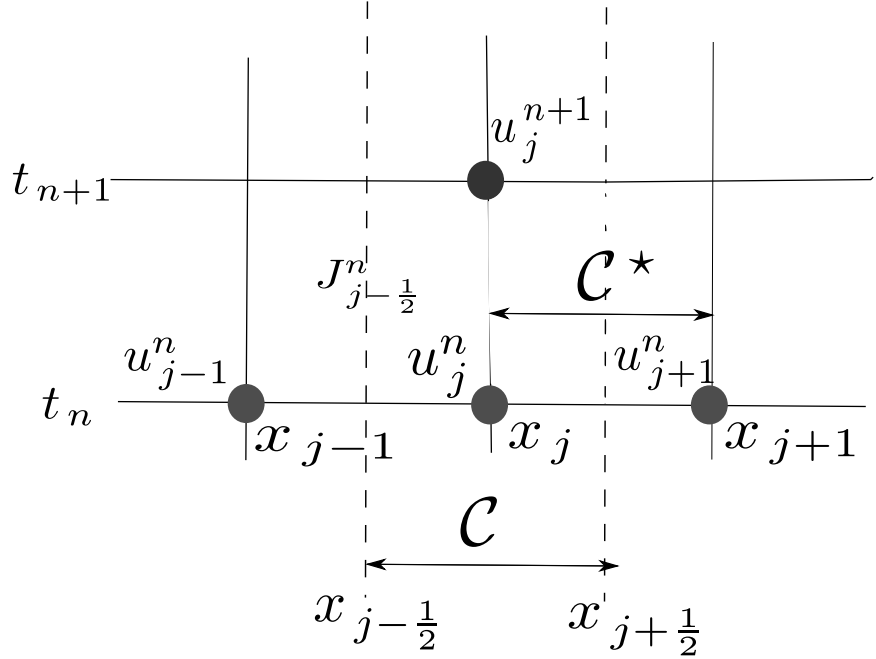}
	\caption{\small\em Stencil of the Scharfetter-Gummel numerical scheme}
	\label{fig:stencil}
\end{figure}

The flux $J$ is denoted by the following expression:
\begin{equation*}
  J \egal d\, \cdot\,\pd{u}{x} \moins a\, \cdot\,u \moins k\,(u) \,.
\end{equation*}
For the cell $\mathcal{C} \egal x \ \in \ \Bigl[\,x_{\,j-\half} \,,\, x_{\,j+\half} \,\Bigr]$ illustrated in Figure~\ref{fig:stencil}, the semi--discretization of the equation~\eqref{eq:main_eq} results as:
\begin{equation}\label{eq:flux_eqn}
  \frac{\ud\,u_{\,j}}{\ud\,t} \egal \nicefrac{1}{\Delta x}\; \bigg(\;J_{\,j+\frac{1}{2}}^{\,n} \moins  J_{\,j-\frac{1}{2}}^{\,n} \;\bigg)\,. 
\end{equation}
Within the \SG ~approach, the assumption is that the flux $J$ is constant on the dual cell $\mathcal{C}^{\,\star} \egal \,\Bigl[\,x_{\,j}\,,\,x_{\,j+1}\,\Bigr]\,$. The following boundary value problem can be written:
\begin{subequations}\label{eq:flux_discr}
\begin{align}
  J_{\,j+\frac{1}{2}}^{\,n} \ &= \ d\, \cdot \,\pd{u}{x} \moins a\, \cdot\,u \moins k^n_{j+\frac{1}{2}} \,, 
  && \forall x \, \in \, [\,x_{\,j},\, x_{\,j+1}\,]\,,\\
  u \ &= \ u^{\,n}_{\,j}\,,
  && x\ = \ x_{\,j} \,,\\
  u \ &= \ u^{\,n}_{\,j+1}\,, 
  && x\ = \ x_{\,j+1} \,.
\end{align}
\end{subequations}
where $k^{\,n}_{\,j+\frac{1}{2}}$ is approximated by:
\begin{align*}
  k^{\,n}_{\,j+\frac{1}{2}} \egal k \, \Biggl(\, \frac{1}{2}\, \Big(\, u^{\,n}_{\,j} \plus u^{\,n}_{\,j+1}\,\Big)\, \Biggr) \,.
\end{align*}
It is important to note that problem~\eqref{eq:flux_discr} has two unknowns $u\,(\,x\,,\,t\,)$ and $J_{\,j+\frac{1}{2}}^{\,n}$ with two constrains at $x \egal x_{\,j}\,$ and $x \egal x_{\,j+1}\,$. Therefore, one can obtain the exact solution of Eq.~\eqref{eq:flux_discr} as:
\begin{align}\label{eq:exact_flux}
  J_{\,j+\frac{1}{2}}^{\,n} \egal \moins a\, \cdot\; \nicefrac{\Bigl(\, u^n_{j+1}\moins u^n_j\, e^{\nicefrac{a \, \Delta x}{d}} \, \Bigr)}{1 \moins e^{\nicefrac{a \, \Delta x}{d}}} \moins k^n_{j+\frac{1}{2}} \,.
\end{align}
Applying expression~\eqref{eq:exact_flux} to Eq.~\eqref{eq:flux_eqn}, and using an \textsc{Euler} explicit approach, we obtain the expression to compute $u_{\,j}^{\,n+1}\,$:
\begin{align*}
  u^{\,n+1}_{\,j} \egal u^n_{j} \plus \frac{\Delta t}{\Delta x}\, \cdot\,\Bigg[\;\moins a\, \cdot\,\nicefrac{\Bigl(\, u^n_{j+1}\plus u^n_{j-1}\, e^{\nicefrac{a \, \Delta x}{d}} \, \Bigr)}{1 \moins e^{\nicefrac{a \, \Delta x}{d}}} \plus\\ 
  a\, \cdot\,\nicefrac{\Bigl(\, 1 \plus e^{\nicefrac{a \, \Delta x}{d}} \, \Bigr)}{1 \moins e^{\nicefrac{a \, \Delta x}{d}}}\, \cdot\,u^n_j \moins k^n_{j+\frac{1}{2}} \plus k^n_{j-\frac{1}{2}}\,\Bigg] \,.
\end{align*}
In this work, several approaches are used for the temporal discretization. As presented above \textsc{Euler} explicit approach can be applied. Additionaly, the solution is computed with adaptive time step the \textsc{Adams}--\textsc{Bashforth}--\textsc{Moulton} algorithm, using \texttt{Matlab\texttrademark} function \texttt{ODE113} \cite{Shampine1997}.

\subsection{Properties of \SG ~scheme}
\label{sec:properties_SG}

The main advantage of the \SG ~numerical scheme is its explicit formulation of the solution. Thus, the nonlinearities are handled without costly sub--iterations at each time steps as it is the case with implicit approaches. Another significant point, the scheme is well--balanced as well as asymptotically preserving \cite{Gosse2017}. The \SG ~scheme has first order accuracy over the time and space $\O\,(\Delta x\ +\ \Delta t)$ \cite{GartlandJr.1993}. Moreover, the flux is also approximated to the first order $\O\,(\,\Delta x\,)\,$.

For Eq.~\eqref{eq:main_eq}, the \textsc{Courant}--\textsc{Friedrichs}--\textsc{Lewy} (\textsc{CFL}) stability condition for the \SG ~numerical scheme combined with an \textsc{Euler} explicit approach, is calculated with the expressions from \cite{Gosse2017}:
\begin{align}\label{eq:CFL_lin}
  \Delta t \, \cdot \, \max_{\, j} \Biggl[\, \biggl(\, a \plus \frac{\mathrm{d}k}{\mathrm{d}u}\, \Bigl(\,u\,\bigl(\,x_{\,j+\half}\,\bigr)\,\Bigr)\,\biggr)\, \cdot\,\tanh\Biggl(\, \biggl(\, a \plus \frac{\mathrm{d}k}{\mathrm{d}u}\, \Bigl(\,u\,\bigl(\,x_{\,j+\half}\,\bigr)\,\Bigr) \,\biggl) \, \cdot\, \nicefrac{\Delta x}{2\,d}\Biggr)^{-1}\,\Biggr]\nonumber \\ 
  \leq \, \Delta x \,.
\end{align}
Through this condition, a nonlinear relation stands between $\Delta t$ and $\Delta x\, $. But if large space discretization $\Delta x$ is taken, time and space grid become proportional to each other $\Delta t \ \simeq \ \Delta x\,$. Thus, the stability condition is relaxed compared to the classical central finite--difference scheme combined with \textsc{Euler} explicit approach $ \Delta t \ \simeq \ \Delta x^{\,2}\,$.

As mentioned before, Eq~\eqref{eq:flux_discr} has two unknowns $u\,(\,x\,,\,t\,)$ and $J_{\,j+\frac{1}{2}}^{\,n}$. The solution $J_{\,j+\frac{1}{2}}^{\,n}$ is given by Eq.~\eqref{eq:flux_discr}. The exact expression of solution $u$ can be computed for $x \, \in \, \mathcal{C}^{\,\star}\,$: 
\begin{align*}
  u^{\,n}(\,x\,) \egal \nicefrac{u^n_j\,\ue^{\nicefrac{a\, \Delta x}{d}} \moins u^n_{j+1}}{\ue^{\nicefrac{a \,\Delta x}{d}} \moins 1} \plus \nicefrac{u^n_{j+1}\moins u^{n}_j}{\ue^{\nicefrac{a \,\Delta x}{d}} \moins 1}\, \cdot\,\ue^{\Bigg({\dfrac{a\, \cdot\,(x \moins x_j)}{d}}\Bigg)}, \\ 
  x \ \in \ \Bigl[\,x_{\,j}\,,\, x_{\,j+1}\,\Bigr] \,. 
\end{align*}


\subsection{Extension for the nonlinear case}
\label{sec:extension_nonlinear_case}

In case where the coefficients $a$, $d$ and $k$ of Eq.~\eqref{eq:main_eq} all depend on $u\,,$ we approximate them by using the frozen coefficients assumption on the dual cell $\Bigl[\,x_{\,j}\,,\, x_{\,j+1} \,\Bigr]\,$.

Thus, the solution of the boundary-value problem~\eqref{eq:flux_discr} is written as:
\begin{align*}
  J_{\,j+\frac{1}{2}}^{\,n} \egal \moins a^{\,n}_{\,j+\frac{1}{2}} \, \cdot\, \nicefrac{\Big(\,u^{\,n}_{\,j+1} \moins u^{\,n}_{\,j} \, \ue^{\theta^{\,n}_{\,j+\frac{1}{2}}} \,\Big)}{1 \moins \ue^{\,\theta^{\,n}_{\,j+\frac{1}{2}}}} \moins k^{\,n}_{\,j+\frac{1}{2}} \,,
\end{align*}
where 
\begin{equation*}
  \theta^{\,n}_{\,j+\frac{1}{2}} \egal \,\,{\nicefrac{a^{\,n}_{\,j+\frac{1}{2}} \, \cdot\, \Delta x}{d^{\,n}_{\,j+\frac{1}{2}}}}\,,
\end{equation*}
and
\begin{align}\label{eq:nonlinear_coefficients}
  & a^{\,n}_{\,j+\frac{1}{2}} \egal a \,\biggl(\, \frac{1}{2} \Bigl(\, u^n_j + u^n_{j+1} \,\Bigr) \,\biggr) \,,
  && d^{\,n}_{\,j+\frac{1}{2}} \egal d \, \biggl(\, \frac{1}{2} \Bigl(\, u^n_j + u^n_{j+1}\,\Bigr) \,\biggr) \,,
  && k^{\,n}_{\,j+\frac{1}{2}} \egal k \, \biggl(\, \frac{1}{2} \Bigl(\, u^n_j + u^n_{j+1}\,\Bigr) \,\biggr) \,.
\end{align}
For the nonlinear case the CFL stability condition for the \SG ~numerical scheme is is calculated with the expressions from \cite{Gosse2017}:
\begin{align*}
  \Delta t\, \cdot\,\max_{\, j} \, d_{\,j}\, \max_{\, j} \Bigg[\,  \nicefrac{p_{\,j+\frac{1}{2}}}{d_{\,j+\frac{1}{2}}} \,\tanh\Bigg(\nicefrac{p_{\,j+\frac{1}{2}}\, \cdot\,\Delta x}{2\,d_{\,j+\frac{1}{2}}}\Bigg)^{-1}\,\Bigg] \,\leq \, \Delta x \, ,
\end{align*}
where
\begin{align*}
  & p_{\,j+\frac{1}{2}} \egal a_{\,j+\frac{1}{2}} \plus u_{\,j+\frac{1}{2}}\, \cdot\,\frac{\mathrm{d}a}{\mathrm{d}u}\, \Bigl(\,u\,\bigl(\,x_{\,j+\half}\,\bigr)\,\Bigr)\, \plus \frac{\mathrm{d}k}{\mathrm{d}u}\, \Bigl(\,u\,\bigl(\,x_{\,j+\half}\,\bigr)\,\Bigr)  \,.
\end{align*}


\subsection{Comparing numerical models}

In order to validate the numerical model, the error between the solution $u^{\,\mathrm{num}}\,(x,\,t)$ and a reference one $u^{\mathrm{ref}}\,(x,\,t)$ is evaluated as a function of $x$ according to the formula:
\begin{align*}
  \varepsilon_{\,2}\,(x) \, \equiv \, \sqrt{\,\nicefrac{1}{N_t} \sum_{j=1}^{N_t} \Bigg(\,u^{\,\mathrm{num}}\big(\,x\,,\,t_{\,j}\,\big)\moins u^{\,\mathrm{ref}}\big(\,x,t_{\,j}\,\big)\,\Bigg)^2\,} \,,
\end{align*}
where $N_{\,t}$ is the number of temporal steps.

The local uniform error is defined as the maximum value of $\varepsilon_{\,2}\,(x)\,$:
\begin{align*}
  \varepsilon_{\, \infty} \, \equiv \, \max_{x \, \in \,[\,0,\,L_{\,x}\,]} \, \varepsilon_{\,2}\,(x) \, .
\end{align*}

The \textit{significant correct digits} (scd) of the solution is calculated according to \cite{Soderlind2006}:
\begin{align*}
  \scd(u) \, \equiv \, - \log_{10} \ \norm*{ \frac{u(\,x,t_{\,\mathrm{end}}\,) \moins  u^{\,\mathrm{ref}}(\,x,t_{\,\mathrm{end}}\,)}{u^{\,\mathrm{ref}}(\,x,t_{\,\mathrm{end}}\,)} }_{\,\infty}\,.
\end{align*}

The reference function $u^{\mathrm{ref}}\,\big(\,x,\,t\,\big)$ is given by a numerical solution of the differential equation based on the \textsc{Chebyshev} polynomial and adaptive spectral methods and obtained using the function \texttt{pde23t} from the \texttt{Matlab\texttrademark} open source package \texttt{Chebfun} \cite{Driscoll2014}.


\subsection{Numerical validation}
\label{sec:numerical_validation}

In this Section, the purpose is to perform numerical computations to validate the proposed model. Thus, only dimensionless quantities are considered with no necessary physical meanings.


\subsubsection{Linear case}

First, we study a case with constant functions  $d\,(\,u\,)$, $a\,(\,u\,)$ and a linear function of $k\,(\,u\,)$:
\begin{align*}
  & d\,(\,u\,) \egal d_{\,0} \egal 0.05 \,, 
  && a\,(\,u\,) \egal a_{\,0} \egal 0.02 \,, 
  &&k\,(\,u\,) \egal k_{\,1} \; u \,,
\end{align*}
with $k_{\,1} \egal 0.5\,$. It should be noted that the dimensionless numbers are set to unity in Eq.~\eqref{eq:water_uptake_dimless}. The following boundary and initial conditions are set:
\begin{align*}
  & u\,(\,0\,,\,t\,) \egal 0.2\,\Bigl(\,1 \moins \cos \,(\,\pi \,t\,)^{\,2} \,\Bigr) \,,
  && u\,(\,1\,,\,t\,) \egal 0.3\,\sin(\,\pi \, t\,)^{\,2} \,,
  && u\,(\,x\,,\,0\,) \egal 0 \,.
\end{align*}
The time domain is defined as $t \ \in \ \bigl[\, 0\,,\,3\,\bigr] \,$. According to Equation~\eqref{eq:CFL_lin}, the \SG ~scheme CFL condition is given in this case by: 
\begin{align*}
  \Delta t \leq  \, \nicefrac{\tanh{\biggl(\,\dfrac{\big(\,a_{\,0} \plus k_{\,1}\,\big)\, \cdot\,\Delta x}{\,2\,d_{\,0}}\,\biggr)}}{a_{\,0} \plus k_{\,1}}\, \cdot\,\Delta x \,.
\end{align*}
First, the problem is solved by implementing an adaptive in time approach, using \texttt{Matlab\texttrademark} function \texttt{ODE113} with absolute and relative tolerances set to $10^{\,-\,4}\,$. The space discretization parameter is set $\Delta x \egal 10^{\,-\,2}\,$. Given the numerical value of the parameter, the CFL condition is $\Delta t \leq 10^{\,-\,3}\,$.

\begin{figure}
\begin{center}
\subfigure[]{\includegraphics[width=0.45\textwidth]{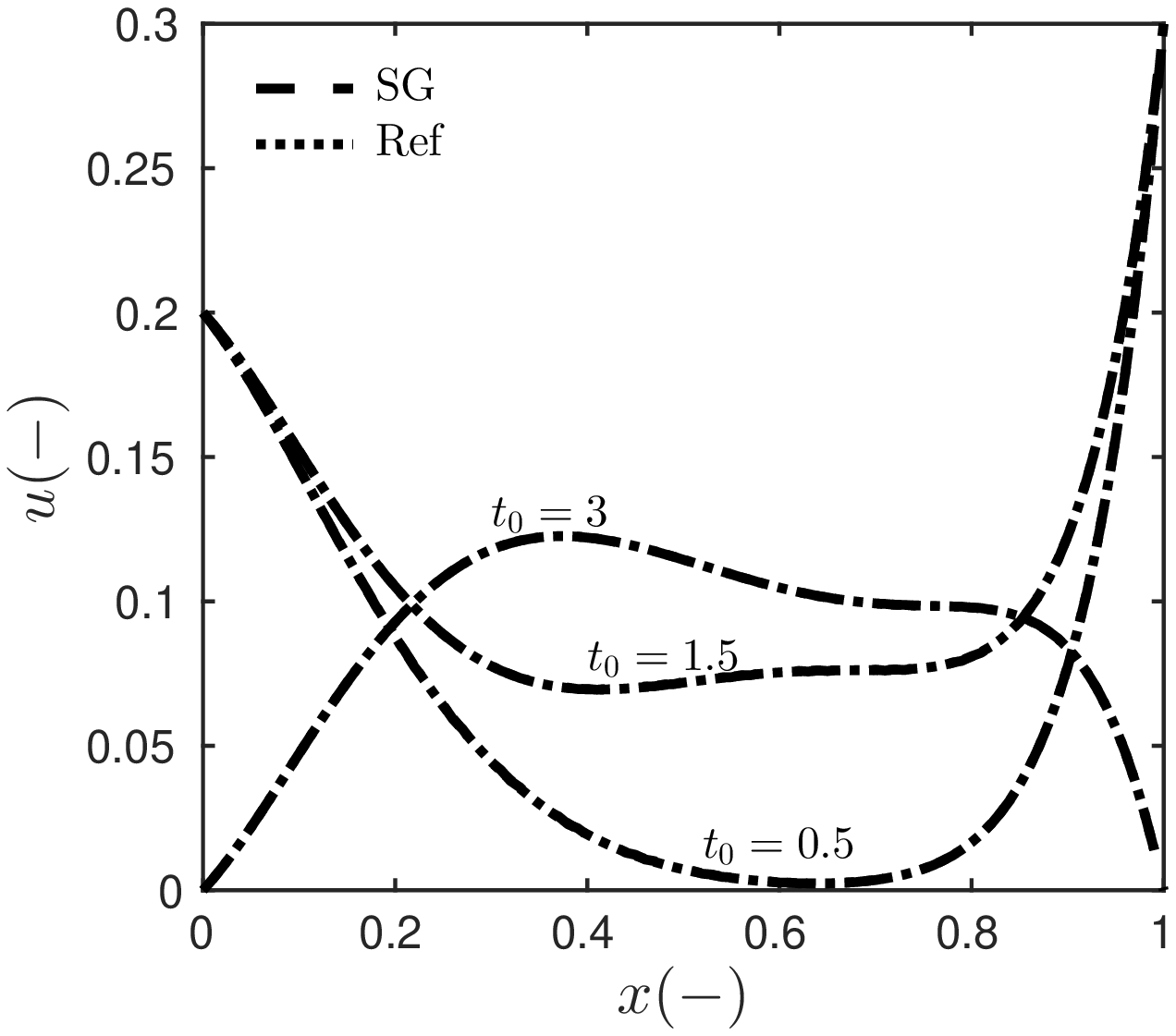}} \hspace{0.3cm}
\subfigure[]{\includegraphics[width=0.45\textwidth]{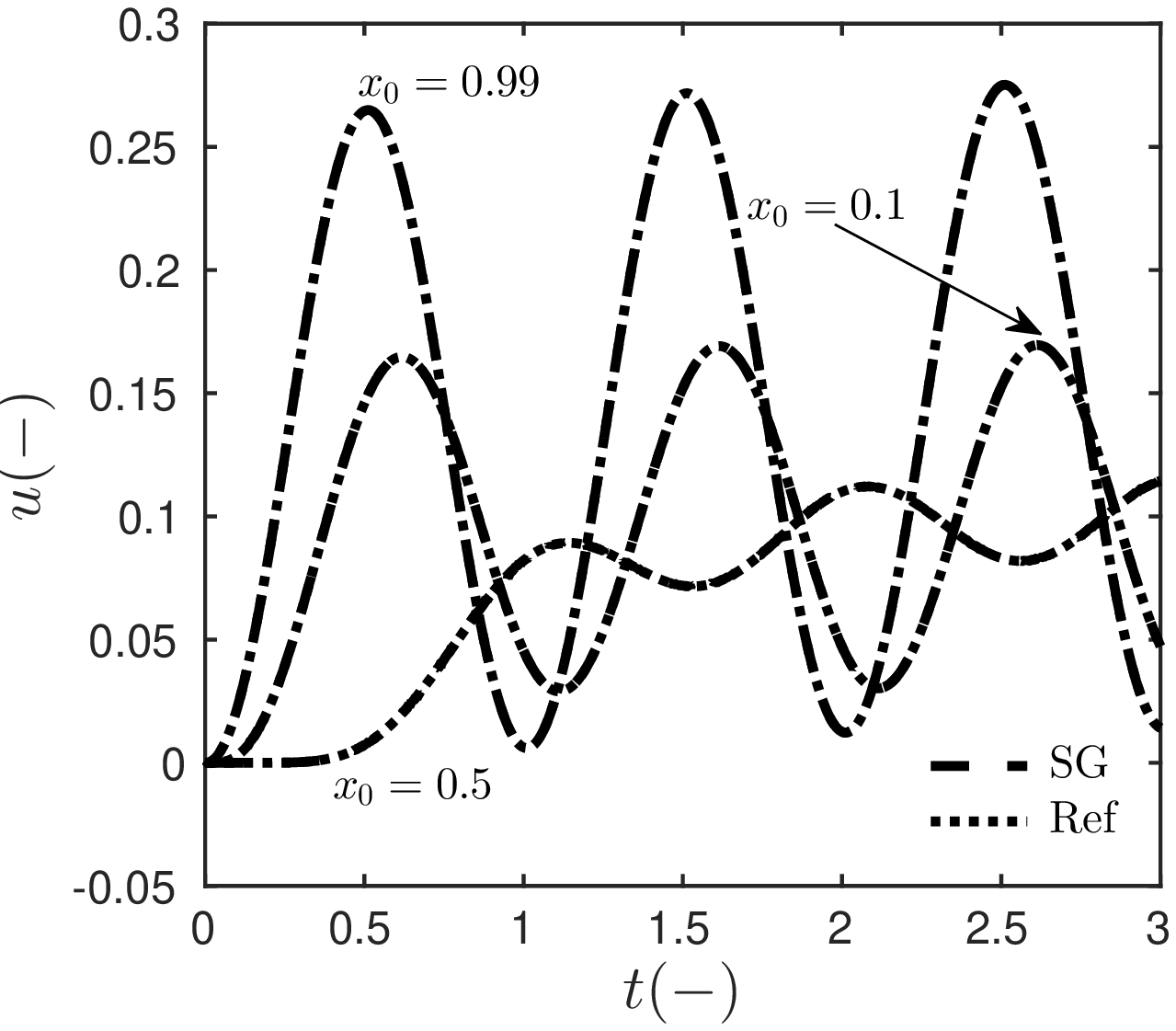}} 
\caption{\small\em Variation of the field $u$ as a function of \emph{(a)} space and \emph{(b)} and time $t\,$.}
\label{fig:solU}
\end{center}
\end{figure}

\begin{figure}
\centering
\includegraphics[width=0.6\linewidth]{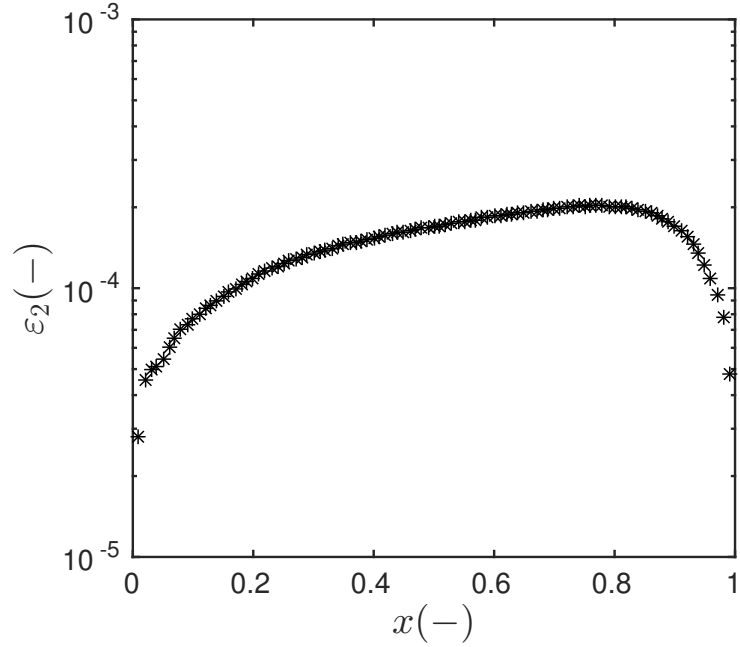}
\caption{\small\em Error $\varepsilon_{\,2}$ as a function of space.}
\label{fig:eps2x}
\end{figure}

Figure~\ref{fig:solU} shows the variations of the solution $u\,(\,x\,,t\,)$ as a function of space and time. It demonstrates a satisfactory agreement between the \SG~numerical solution and the reference one. Figure~\ref{fig:eps2x} highlights the $\mathcal{L}_{\,2}$ error value $\varepsilon_{\,2}\,(\,x\,)$ at the order of $10^{\,-\,4}\,$, which validates the numerical model for this case. This result is consistent with the value of the discretization parameter and the tolerances set in the \texttt{ODE} solver.

\begin{figure}
\begin{center}
\subfigure[\label{fig:eps_dx}]{\includegraphics[width=0.85\textwidth]{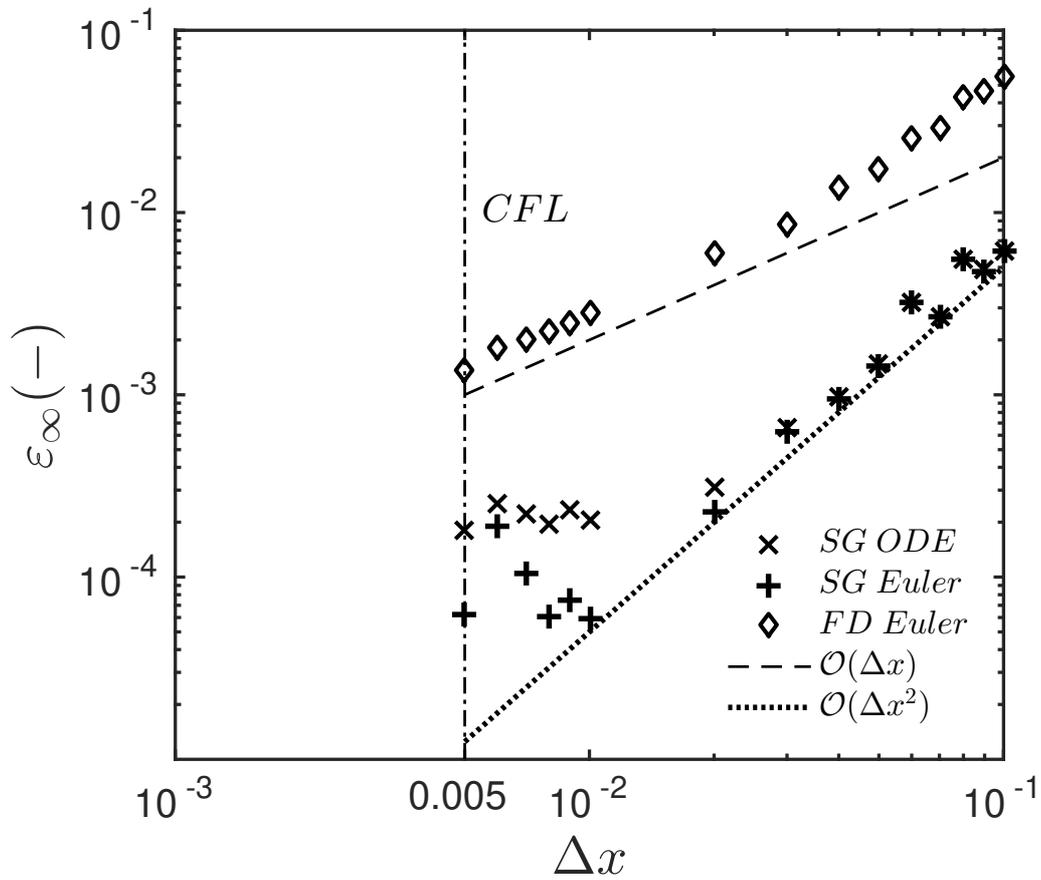}}\\
\subfigure[\label{fig:cpu_dx}]{\includegraphics[width=0.45\textwidth]{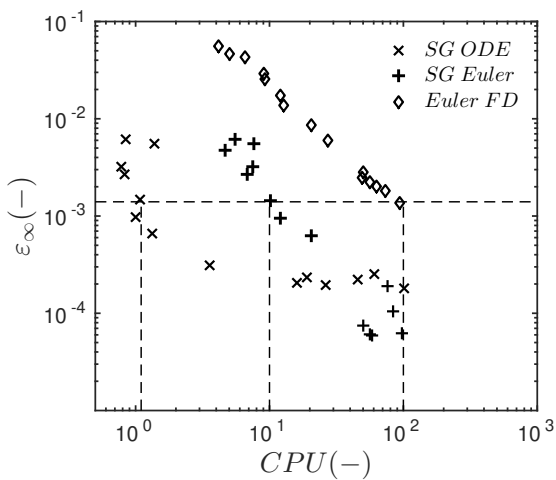}} 
\hspace{0.3cm}
\subfigure[\label{fig:scd_dx}]{\includegraphics[width=0.45\textwidth]{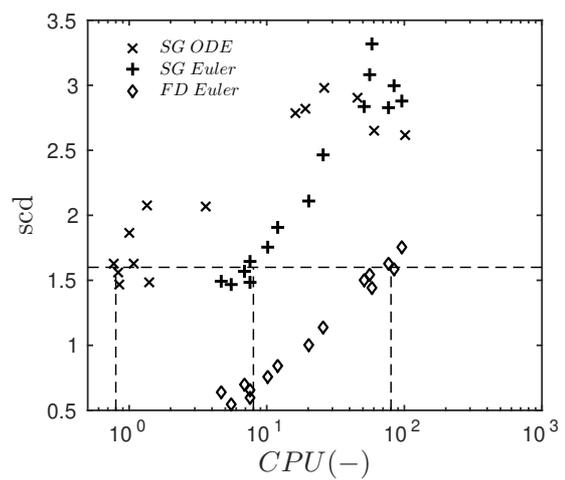}}
\caption{\small\em (a) Variation of the error $\varepsilon_{\,\infty}$  as a function of $\Delta x\,$. (b) Variation of the error $\varepsilon_{\,\infty}$ as a function of the CPU time of the numerical model. (c) Variation of the accuracy digits value $\mathrm{scd}$ as a function of the CPU time of the numerical model.}
\end{center}
\end{figure}

\begin{figure}
\begin{center}
  \includegraphics[width=0.75\textwidth]{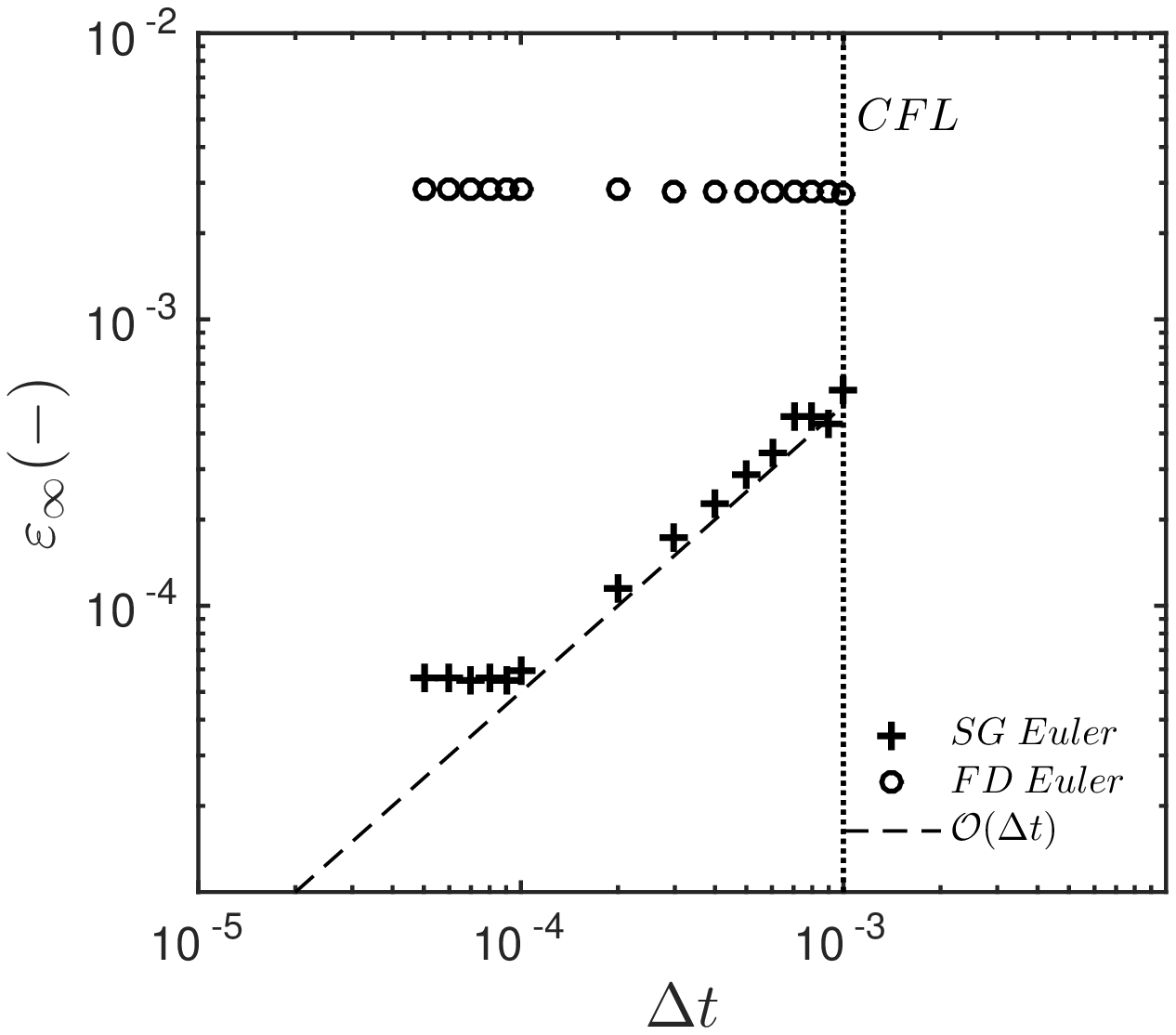}
  \caption{\small\em Variation of the error $\varepsilon_{\,\infty}$  as a function of $\Delta t\,$.}
  \label{fig:eps_dt}
\end{center}
\end{figure}

An analysis of the numerical error is performed by varying one of the discretization parameters $\Delta x$ and $\Delta t$ values, while the other parameter remains constant. Different numerical schemes are compared for the computation of the solution of problem~\eqref{eq:main_eq}. The \SG ~scheme is implemented with (\textit{i}) the adaptive time step approach using \texttt{ODE113} solver and a tolerance set to $10^{\,-\,4}$ and (\textit{ii}) the \Eu ~explicit approach. For comparison, a central finite--difference scheme using the \Eu ~explicit method is also used.

Figure~\ref{fig:eps_dx} shows the variation of the error $\varepsilon_{\,2}$ as a function of $\Delta x$ with a fixed time discretization parameter $\Delta t \egal 10^{\,-4}$. The CFL condition $\Delta t \ \leqslant \ 2.5 \cdot 10^{\,-4}$ is respected until  $\Delta x \egal 5 \cdot 10^{\,-3}\,$. Beyond this limit, the numerical model cannot provide a bounded solution. These results also confirm a higher accuracy of the numerical model using the \SG ~scheme. Moreover, the \SG ~numerical scheme seems to have a second-order accuracy in space $\mathcal{O}(\,\Delta x^{\,2}\,)\,$ which is better than the theoretical results mentioned in Section~\ref{sec:properties_SG}. The variations of the computational time with the requested level of accuracy is shown in Figure~\ref{fig:cpu_dx}. For the accuracy $\varepsilon_{\,\infty} \egal \mathcal{O}\,(\,10^{-3}\,)\,,$ the finite--difference is ten-time slower than the \SG ~approach combined with the \Eu ~explicit method. It can be noted that the \SG ~scheme with the adaptive time step is the fastest one thanks to the nonlinearity of the stability condition and adaptive time step. The computational time can be reduced by a hundred times. Figure~\ref{fig:scd_dx} displays the variation of the significant correct digits with the computational time. In order to reach $\mathrm{scd} \egal \O(\,1.5\,)\, $ the finite--difference scheme requires ten times larger CPU time than the \SG ~scheme with \Eu ~explicit approach. The latter is ten times slower than the one with an adaptive approach.

Figure~\ref{fig:eps_dt} presents another error study conducted by varying $\Delta t$ while the space discretization is fixed to $\Delta x \egal 10^{\,-2}\,$. For the given parameters, the CFL condition is $\Delta t \ \leqslant \ 10^{\,-3}\,$. As noticed in Figure~\ref{fig:eps_dt}, a bounded solution can be computed until this condition is respected. It can also be noted that the \SG ~scheme with \Eu ~explicit approach has a greater degree of accuracy. Moreover, the scheme is first--order accurate in time $\mathcal{O}\,(\,t\,)\,$ as mentioned by the theoretical results.


\subsubsection{Nonlinear case}

The previous validation is performed for a linear case in order to verify the theoretical results and advantages of the \SG ~numerical scheme. Since, the parameter estimation problem to be solved in Section~\ref{sec:parameter_estimation} considers nonlinear coefficients, this Section aims at validating the numerical model for such cases. We consider the problem~\eqref{eq:water_uptake_dimless} with the following nonlinear coefficients:
\begin{align*}
  & a\,(\,u\,) \egal 0.02 \plus 0.01\,u^{\,2}  \,, 
  && d\,(\,u\,) \egal 0.05 \plus 0.03\,u^{\,2}  \,, 
  && k\,(\,u\,) \egal 0.05\,u^{\,2} \,.
\end{align*}
The dimensionless numbers are set to unity. The initial and boundary conditions are:
\begin{align*}
  & u\,(\,x\,,\,0\,) \egal 0 \,, \\[4pt]
  & u\,(\,0\,,t\,) \egal 0.8\,\sin\,\Bigl(\,{\pi\,t}\,/\,{3}\,\Bigr) \plus 0.2\,\sin\,\Bigl(\,{\pi\,t}\,/\,{5}\,\Bigr)\,, \\[4pt]
  & u\,(\,1\,,t\,) \egal 0.5\,\sin\,\Bigl(\,{\pi\,t}\,/\,{4}\,\Bigr) \plus 0.3\,\sin\,\Bigl(\,{\pi\,t}\,/\,{7}\,\Bigr) \,.
\end{align*}
The solution of this problem is computed for the time horizon $t \, \in \, \bigl[\, 0 \,,\, 15\,\bigr]\,$. The \SG ~numerical scheme is used with a space discretization step $\,\Delta\,x \egal 0.01\,$ and an adaptive time step with error tolerances set to $10^{\,-\,4}\,$.

\begin{figure}
\begin{center}
\subfigure[]{\includegraphics[width=0.45\textwidth]{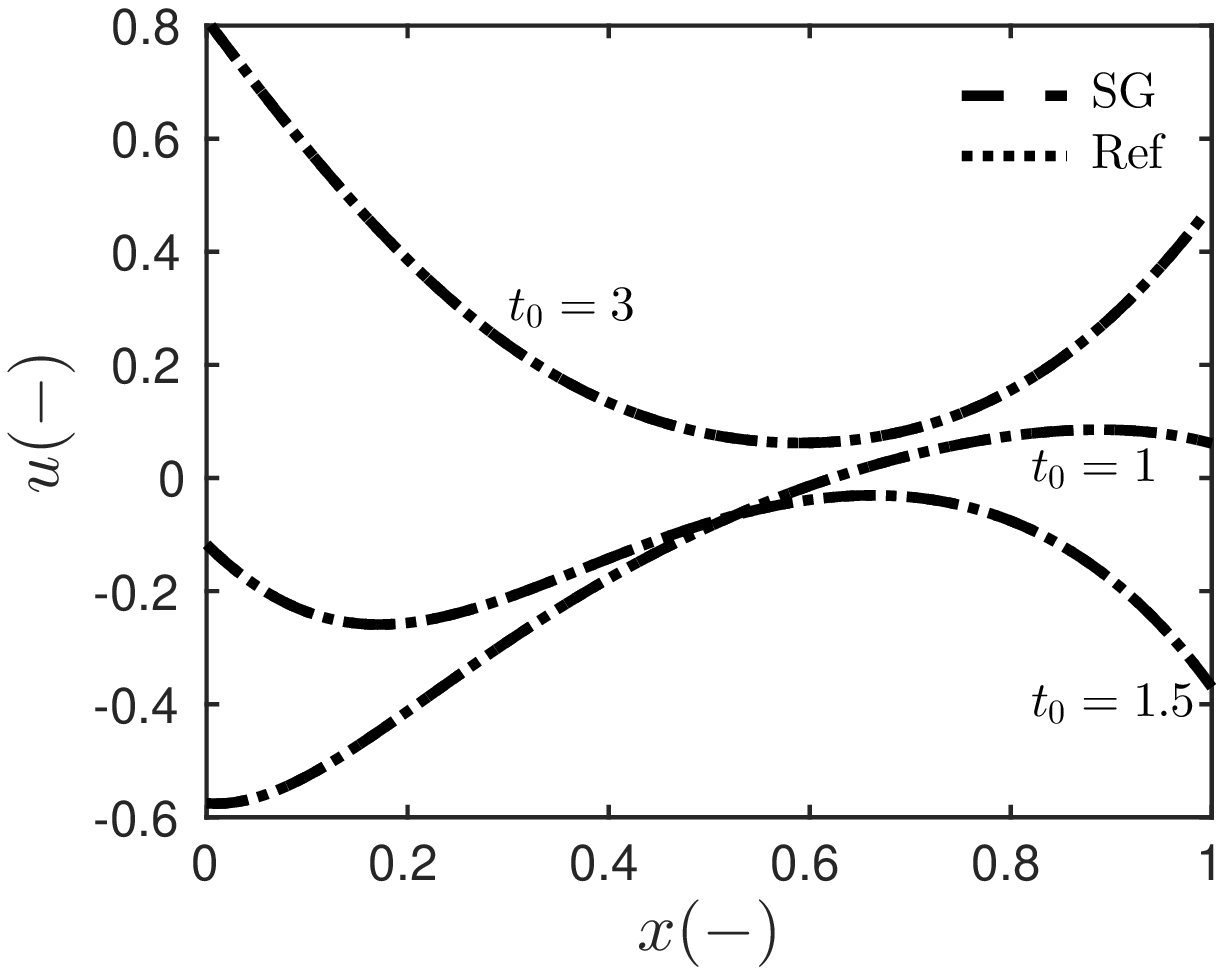}} \hspace{0.3cm}
\subfigure[]{\includegraphics[width=0.45\textwidth]{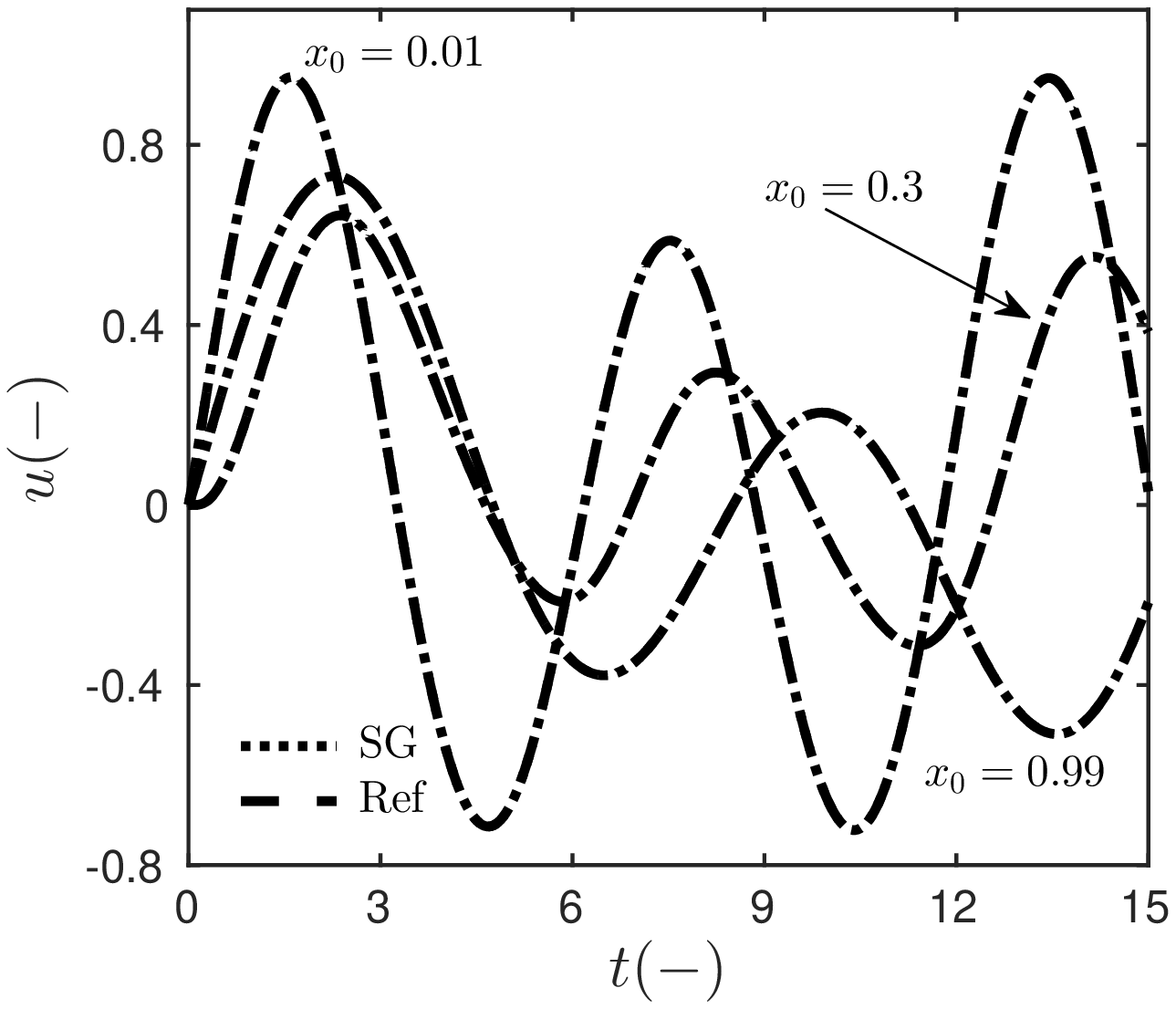}} 
\caption{\small\em Variation of the field $u$ as a function of $x\;$\emph{(a)} and $t\;$\emph{(b)} with nonlinear coefficients}
\label{fig:u_nonlin}
\end{center}
\end{figure}

As shown in Figure~\ref{fig:u_nonlin}, the profiles of the solution shows a very satisfactory agreement between the \SG ~solution and the reference one. It validates the proposed numerical model. The computational time of the numerical model is compared to three other ones: \textit{(i)} the \SG ~scheme with \Eu ~explicit approach, \textit{(ii)}  the central finite--differences approach with \Eu ~explicit and \textit{(iii)}  the central finite--differences with adapting time step. As noticed in Table~\ref{table:cpu_nonlin}, the \SG ~with the adaptive time step scheme is the faster than \SG ~with \Eu ~explicit approach. The computational time is of the same order between the two adaptive in time approaches (\SG ~and central finite--differences). However, the \SG ~with adaptive time step approach is ten times more accurate than central finite--difference with adaptive time step.

\begin{table}
\centering
\caption{\small\em Computational time required to compute the solution of the nonlinear case with the different numerical models.}
\bigskip
\begin{tabular}{c c c c} 
\hline
\hline
{\textbf{Numerical model}}  & \textbf{CPU time $[\,\mathsf{s}\,]$} & \textbf{CPU time $[\,\mathsf{\%}\,]$} & $\boldsymbol{\varepsilon}_{\,\infty}$\\ \hline
SG~with adaptive time step & $105$ & $42$ & $1.9\cdot 10^{-4}$ \\
SG~with \Eu ~explicit & $249$ & $100$ & $5.8\cdot 10^{-5}$ \\
Central FD with adaptive time step & $101$ & $40$ & $1.1\cdot 10^{-3}$ \\
Central FD with \Eu ~explicit & $230$ & $92$ & $2.1\cdot 10^{-2}$ \\
\hline
\hline
\end{tabular}
\label{table:cpu_nonlin}
\end{table}

As mentioned in Section~\ref{sec:extension_nonlinear_case}, the numerical model requires the computation of the coefficients $\,d_{\,j+\half}\,$, $\,a_{\,j+\half}\,$ and $\,k_{\,j+\half}\,$ at the interface of the dual cell $\,\bigl[\,x_{\,j} \,,\, x_{\,j+1}\,\bigr]\,$. According to Eq.~\eqref{eq:nonlinear_coefficients}, these coefficients are interpolated using the mean values of $\,u_{\,j}\,$ and $\,u_{\,j+1}\,$. The accuracy of this interpolation is tested by using different expressions in the computation of the solution $\,u\,$. The coefficients $\,d_{\,j+\half}\,$, $\,a_{\,j+\half}\,$ and $\,k_{\,j+\half}\,$ are calculated by one the following expressions:
\begin{subequations}\label{eq:interp_exp}
\begin{align}
\,k_{\,j+\half} &\egal k \, \Bigg(\, \nicefrac{u_{\,j}\plus u_{\,j+1}}{2} \,\Bigg) \,,\\[3pt]
\,k_{\,j+\half}  &\egal \nicefrac{1}{2} \,\biggl(\, \, k \, \bigl(\, u_{\,j} \,\bigr) 
\plus k \, \bigl(\, u_{\,j+1} \,\bigr) \, \biggr) \,,\\[3pt]
\,k_{\,j+\half} & \egal \frac{2 \, k \, \bigl(\, u_{\,j} \,\bigr)\, \cdot \, k \, \bigl(\, u_{\,j+1} \,\bigr) }{k \, \bigl(\, u_{\,j} \,\bigr) \plus k \, \bigl(\, u_{\,j+1} \,\bigr)} \,,\\[3pt]
\,k_{\,j+\half} & \egal \frac{1}{3} \,\biggl(\, \, k \, \bigl(\, u_{\,j} \,\bigr) 
\plus \sqrt{k \, \bigl(\, u_{\,j} \,\bigr) \, \cdot\, k \, \bigl(\, u_{\,j+1} \,\bigr)}
\plus k \, \bigl(\, u_{\,j+1} \,\bigr) \, \biggr) \,,\\[3pt]
\,k_{\,j+\half} & \egal \sqrt{k \, \bigl(\, u_{\,j} \,\bigr) \, \cdot \, k \, \bigl(\, u_{\,j+1} \,\bigr)} \,, \\[3pt]
\,k_{\,j+\half} & \egal \Biggl(\, \frac{1}{2} \,\biggl(\, \, k \, \bigl(\, u_{\,j} \,\bigr)^{\,3} 
\plus k \, \bigl(\, u_{\,j+1} \,\bigr)^{\,3} \, \biggr) \, \Biggr)^{\frac{1}{3}} \,,\\[3pt]
\,k_{\,j+\half} & \egal \sqrt{ \frac{1}{2} \,\biggl(\, \, k \, \bigl(\, u_{\,j} \,\bigr)^{\,2} 
\plus k \, \bigl(\, u_{\,j+1} \,\bigr)^{\,2} \, \biggr) \, } \,.
\end{align}
\end{subequations}

\begin{figure}
\begin{center}
\subfigure[]{\includegraphics[width=0.45\textwidth]{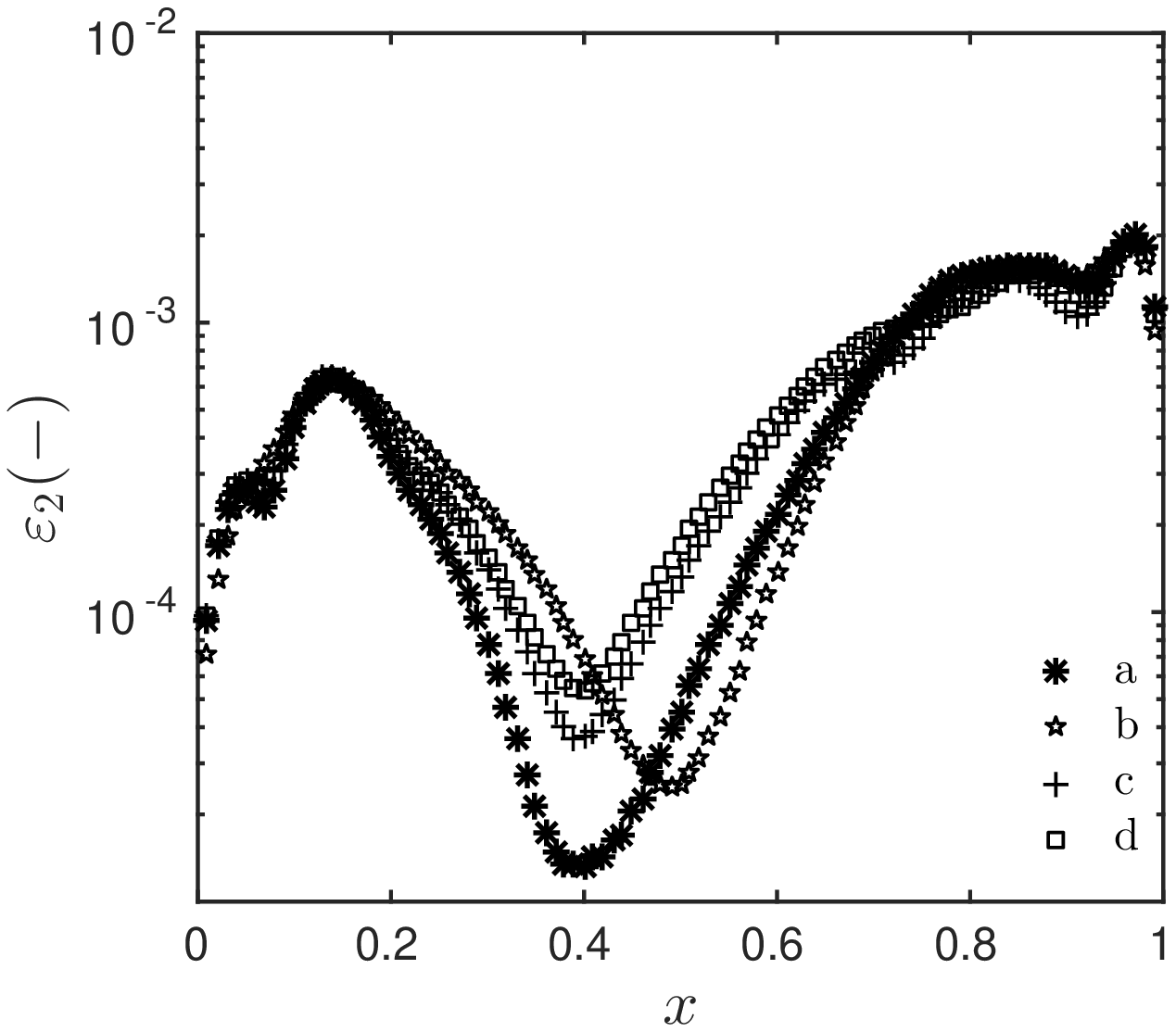}} \hspace{0.3cm}
\subfigure[]{\includegraphics[width=0.45\textwidth]{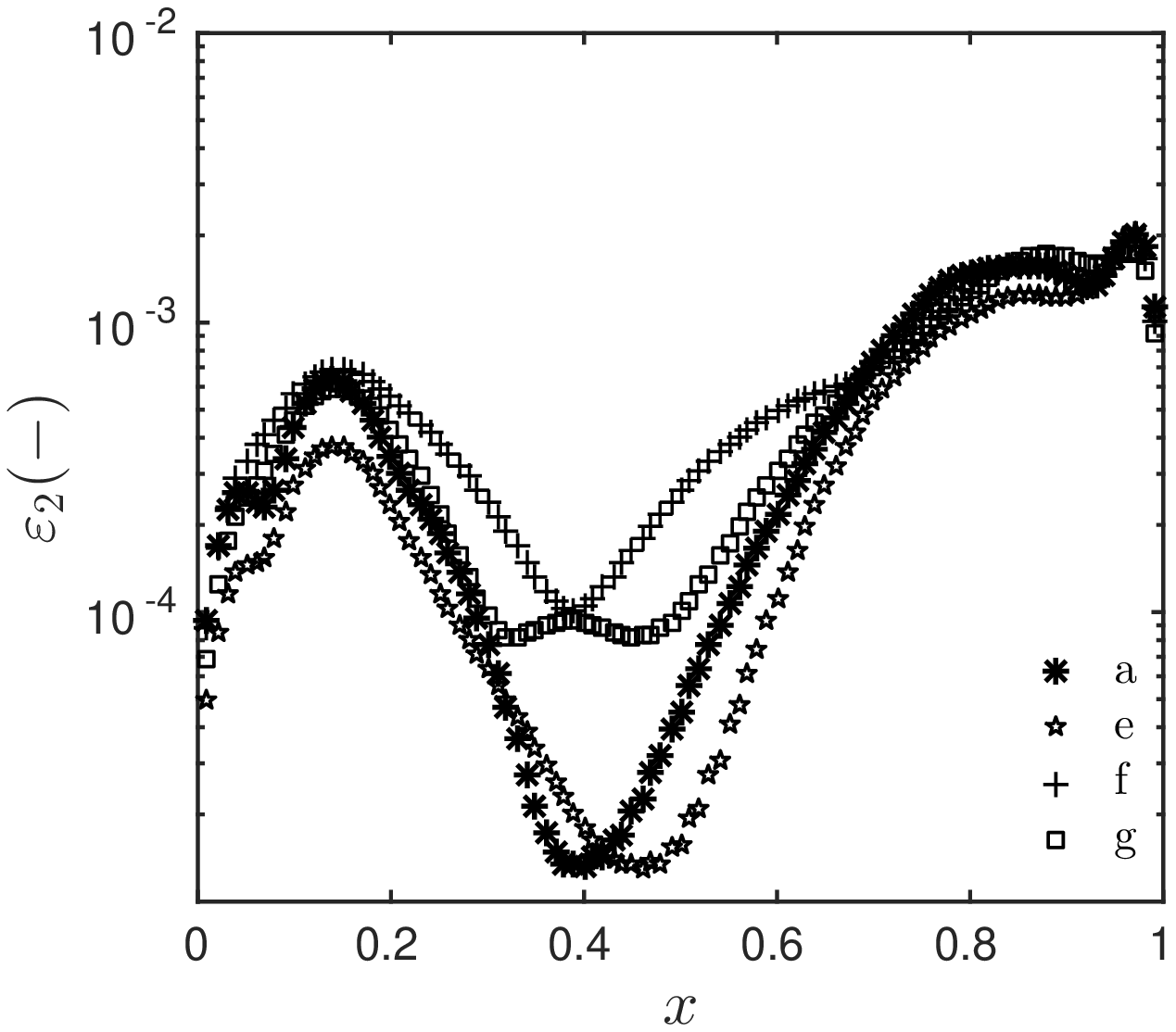}} 
\caption{\small\em Variation of the error $\varepsilon_{\,2}$ with space according to the interpolation formulas Eq.~\eqref{eq:interp_exp} of the coefficients $d\,(\,u\,)\,$, $a\,(\,u\,)$ and $k\,(\,u\,)$.}
\label{fig:coef_interp}
\end{center}
\end{figure}

For each interpolation expression, the CPU time to compute the solution and the error with the reference solution are analyzed. The results are synthesized in Table~\ref{table:interp} and Figure~\ref{fig:coef_interp}. All interpolation expressions have a similar order of accuracy. Nevertheless, it highlights that the approach using the arithmetic mean between $u_{\,j}$ and $u_{\,j+1}\,$ to compute the coefficients $\,d_{\,j+\half}\,$, $\,a_{\,j+\half}\,$ and $\,k_{\,j+\half}\,$ is slightly faster for this particular case.

\begin{table}
\centering
\caption{\small\em Variation of the error of the solution and of the CPU time of the numerical models according to the interpolation formulas Eq.~\eqref{eq:interp_exp} of the coefficients $d\,(\,u\,)\,$, $a\,(\,u\,)$ and $k\,(\,u\,)\,$.}
\bigskip
\begin{tabular}{c c c c c c c c} 
\hline
\hline
\textbf{Formula}  & (a) &(b)&(c)&(d)&(f)&(e)&(g) \\ \hline 
$\boldsymbol{\varepsilon}_{\,\infty}$ & $2\,\cdot\,10^{-3}$ & $1.9\,\cdot\,10^{-3}$ & $2\,\cdot\,10^{-3}$ & $1.9\,\cdot\,10^{-3}$ & $2\,\cdot\,10^{-3}$ & $1.9\,\cdot\,10^{-3}$ & $1.7\,\cdot\,10^{-3}$ \\ \hline
 \textbf{CPU time $[\,\mathsf{s}\,]$} & 21.29 & 25.47 & 33.77 & 25.44  & 34.52 & 27.48  & 25.79  \\ \hline 
\textbf{Percentage $[\,\%\,]$} & $61$  & $74$ & $98$ & $74$ & $100$ & $80$ & $75$\\
\hline 
\hline
\end{tabular}  
\label{table:interp}
\end{table}

This section examined and analyzed the \SG ~scheme applying to the diffusion--advection equation with linear and nonlinear coefficients. These results demonstrate a significant reduction of the computational cost without losing the accuracy while using the \SG ~scheme with an adaptive time step approach. Therefore, in further sections, the solutions are computed with the \SG ~combined with an adaptive time step approach. Since the numerical model is validated, the next section introduces the experimental facility employed to obtain the observations used in the parameter estimation problem.


\section{Experimental facility}
\label{sec:exp_facility}

This section presents the facility to produce the experimental observations that will be compared to the numerical predictions of the model to compare its reliability to represent the physical phenomena.


\subsection{Experimental observations}

\begin{figure}
\begin{center}
\subfigure[]{\includegraphics[width=0.48\textwidth]{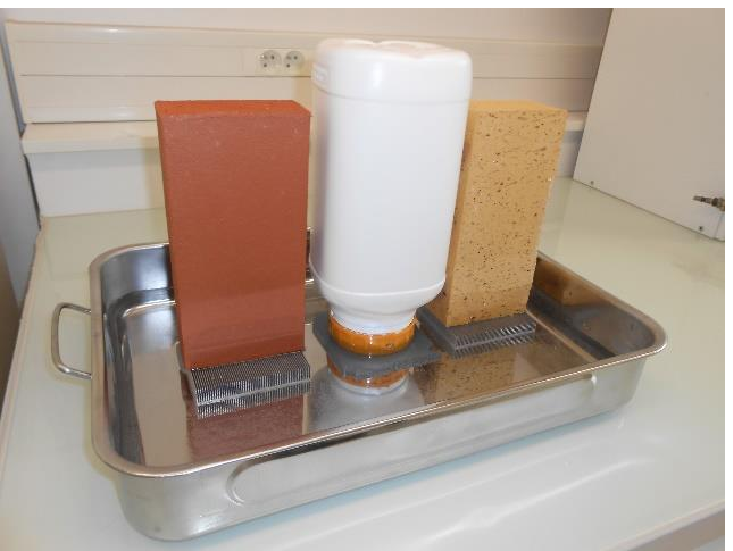}}
\subfigure[]{\includegraphics[width=0.4\textwidth]{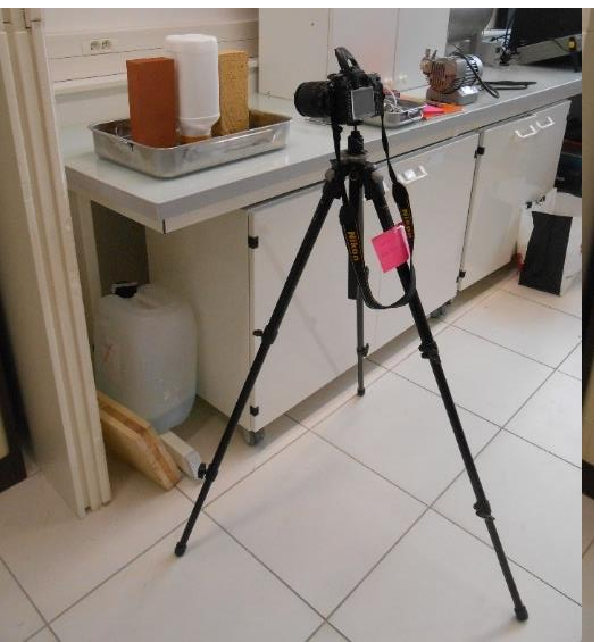}} 
\caption{\small\em Illustrations of the experimental facility. It should be noted that the opaque box has been removed to take the pictures and that the light brick is not the one considered for the present studies.}
\label{fig:exp1}
\end{center}
\end{figure}

To perform the liquid water uptake tests, a refractory brick measuring $11 \, \times \,  22 \, \times \, 5  \ \mathsf{cm}$ is used. The brick is initially sanded allowing a better visualization of the rise of the water. The vertical faces are then taped to be airtight and watertight. As illustrated in Figure~\ref{fig:exp1}, a tray is filled with distilled water and maintained at a constant level thanks to a bottle returned with a pierced cap. The brick is placed vertically on a support at the surface of the water. The experiment begins at the moment when the underside of the brick is in contact with the water and lasts at least $7 \ \mathsf{h} \,$. The rising damp is observed on one of the faces $11 \, \times \,  22 \ \mathsf{cm}$. For this, a camera is placed at $1 \ \mathsf{m}$ to take a picture of the brick every $3 \ \mathsf{min}\,$. The experiment occurs in a box opaque to the light to control the luminosity as well as to minimize the convective and radiative exchanges. Some LED lamps provide a constant lighting in the box.

\begin{figure}
\begin{center}
\subfigure[\label{fig:exp_dp}]{\includegraphics[height=0.45\textwidth]{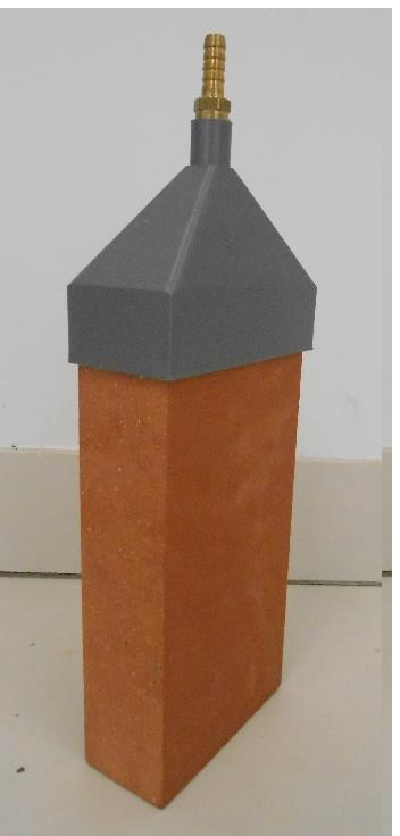}}
\subfigure[\label{fig:exp_grap_paper}]{\includegraphics[width=0.48\textwidth]{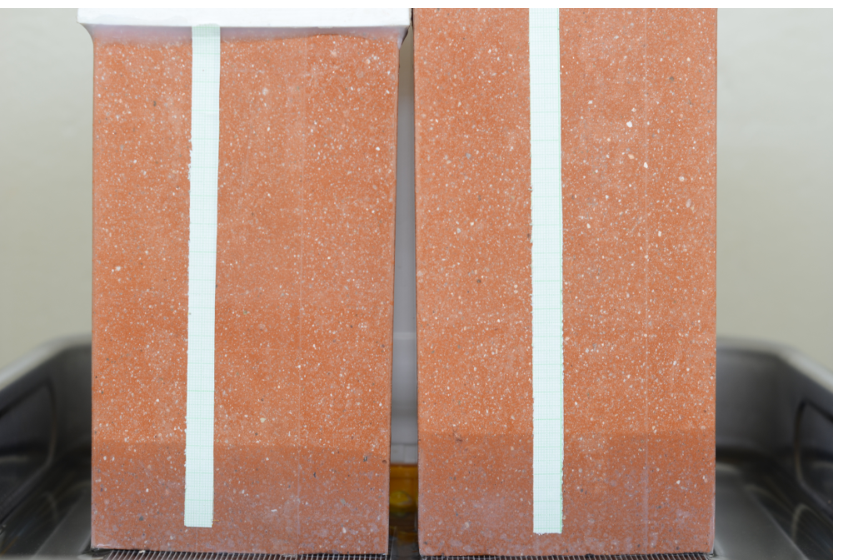}} 
\caption{\small\em (a) Illustrations of the brick submitted to a relative pressure of $- \, 50 \ \mathsf{Pa}$ linking the top of the brick to a fan. (b) Measurement of the water front height using the graph paper.}
\end{center}
\end{figure}

At the top of the brick, two different boundary conditions are imposed. For the first set of experiments, the top of the brick is in contact with the open air at atmospheric pressure. For the second set, a relative pressure of $- \, 50 \ \mathsf{Pa}$ is applied thanks to a chamber with a fan. An illustration of the design is shown in Figure~\ref{fig:exp_dp}.  With this device, the air velocity in the brick is assumed as constant in time. The experiment is taken at the constant room temperature $\;T \egal 293\,\mathrm{K}$.

The height of the water front is taken in the middle of the brick from the pictures of each experiment every $30 \ \mathsf{min}\,$. As shown in Figure~\ref{fig:exp_grap_paper}, graph paper is attached to the brick in order to measure the height front. With this setting, the uncertainty measurement of the height scales with $\sigma_{\,h} \egal 0.5 \ \mathsf{cm}\,$.

\begin{figure}
\centering
\includegraphics[width=0.5\textwidth]{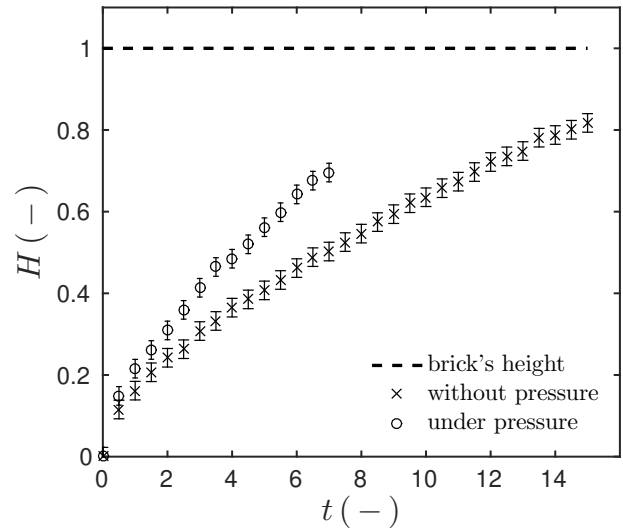}
\caption{\small\em Obtained experimental values of water height for cases without and under pressure. Values are dimensionless, using reference parameters (length of the brick $L$ and $t_{\mathrm{ref}}$)}
\label{fig:exp_data}
\end{figure}

For the case when the brick is exposed to ambient air pressure, the experiment takes $\;t_{\,\mathrm{max}} \egal 15\; \mathsf{h}\,$. For the other experiment, with a pressure difference, it lasts $\;t_{\,\mathrm{max}} \egal 7\; \mathsf{h}\,$. Figure~\ref{fig:exp_data} shows the obtained experimental data for both cases.

\begin{figure}
\centering
\includegraphics[width=0.5\textwidth]{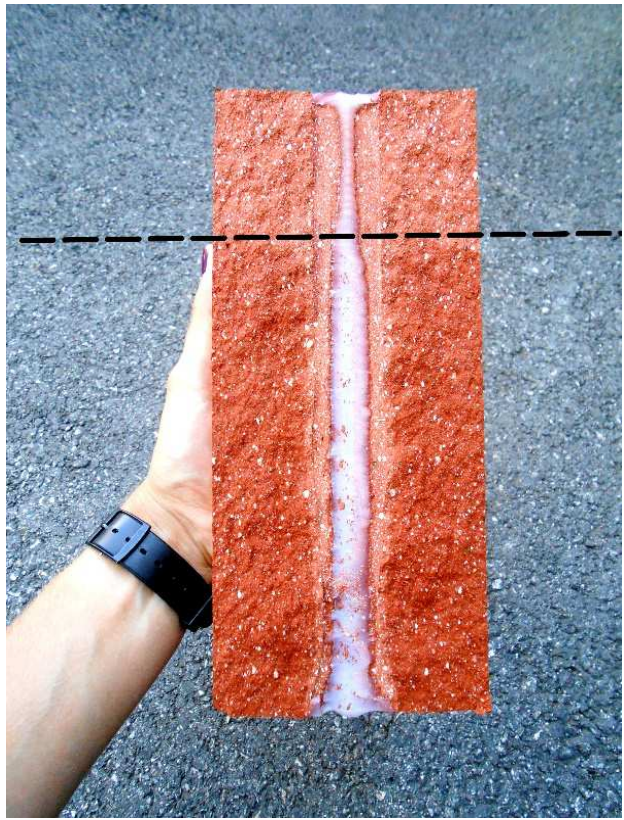}
\caption{\small\em Verification of the uniformity of liquid front inside the brick.}
\label{fig:broken_brick}
\end{figure}

To verify that the liquid front is homogeneous inside the brick and not occurring only at the interfaces where the pictures are taken, a complementary experiment is performed. The brick is preliminary cut in the middle and then fulfilled with silicone to avoid the liquid uptake in this area. The liquid uptake experiment is performed. At the end of the experiments, the cut enables to rapidly break the brick without perturbing the liquid front. As illustrated in Figure~\ref{fig:broken_brick}, it is confirmed that the liquid front is homogeneous inside the brick. The white silicone that fulfills the cut in the middle of the brick can be noticed.

\subsection{Defining the material properties}
\label{sec:parameterization_material properties}

The physical model representing the liquid water uptake test is described in Section~\ref{seq:phys_model}. Now, it is important to define the material properties involved in the definition of the diffusion, gravity and advection fluxes for the brick used in the experiments. These properties are uncertain \cite{Guizzardi2016} and will be determined in the next Section using the experimental data obtained with the facility.

Data from the literature are used to define the dependency of these properties on the moisture content. The data is projected on the polynomial functions, using the \texttt{cftool} from the \texttt{Matlab\texttrademark} environment, and polynomials with the highest root mean square error $\mathbf{R-square}$ (the square of the correlation between the response values and the predicted response values) are selected. These measurements provide \emph{a priori} values of the coefficients of the polynomials, which act as initial guess when solving the parameter estimation problem. The polynomial functions have been chosen for the material properties to fit the data from the literature. Another set of functions can be applied, and the same steps of parameter estimation are utilized. It should be noticed that data fitting is performed only for the liquid water content.

The reference parameters used for computations are $L \egal 0.22\; \mathsf{m}\,$, $t_{\mathrm{ref}} \egal 1\; \mathsf{h}\,$, $D_{\,\mathrm{ref}} \egal  10^{\,-6} \ \mathsf{m^{\,2}/s} \,$, $k_{\,\mathrm{ref}} \egal  10^{\,-10} \ \mathsf{m/s}$ and $a_{\,\mathrm{ref}} \egal 10^{\,-9} \ \mathsf{m/s} \,$. The saturated volumetric moisture content $\,\theta_{\,\mathrm{sat}} \,$, introduced in \cite{KumarKumaran1996}, equals  $\theta_{\mathrm{sat}} \egal 0.3065 \ \mathsf{m^{\,3}/m^{\,3}}$. It induces the following dimensionless numbers:
\begin{align*}
  & \mathrm{Fo} \egal 0.074  \,,
  && \mathrm{Pe} \egal 2.2 \cdot 10^{\,-4} \,,
  && \mathrm{Bo} \egal 0.7 \,.
\end{align*}
With given reference values, dimensionless representation  $d\,(\,u\,)\,$ of the liquid transport coefficient is:   
\begin{align*}
  d\,(\,u\,) \egal d_{\,4} \, u^{\,4} \plus  d_{\,1} \, u \plus d_{\,0}\,, 
\end{align*}
where $d_{\,4} \egal 0.6 \,$, $d_{\,1} \egal -0.04$ and $d_{\,0} \egal 0.0067\,.$ The dimensionless function $a\,(\,u\,)$ of the advection part:
\begin{align*}
  a\,(\,u\,) \egal a_{\,0} \, \Bigl(\,1 \moins H \,\Bigr)\,,
\end{align*}
where $a_{\,0} \egal 0.7 \,,$ and $H \egal \int_{\,0}^{\,1} \, u\,(\,x\,,t\,)\, \mathrm{d}x \,.$  Last, the dimensionless function $k\,(\,u\,)$ of the liquid conductivity part:
\begin{align*}
  k\,(\,u\,) \egal k_{\,3} \, u^{\,3}\,,
\end{align*}
where $ k_{\,3} \egal 0.8\,.$

\begin{figure}
\centering
\includegraphics[width=0.7\textwidth]{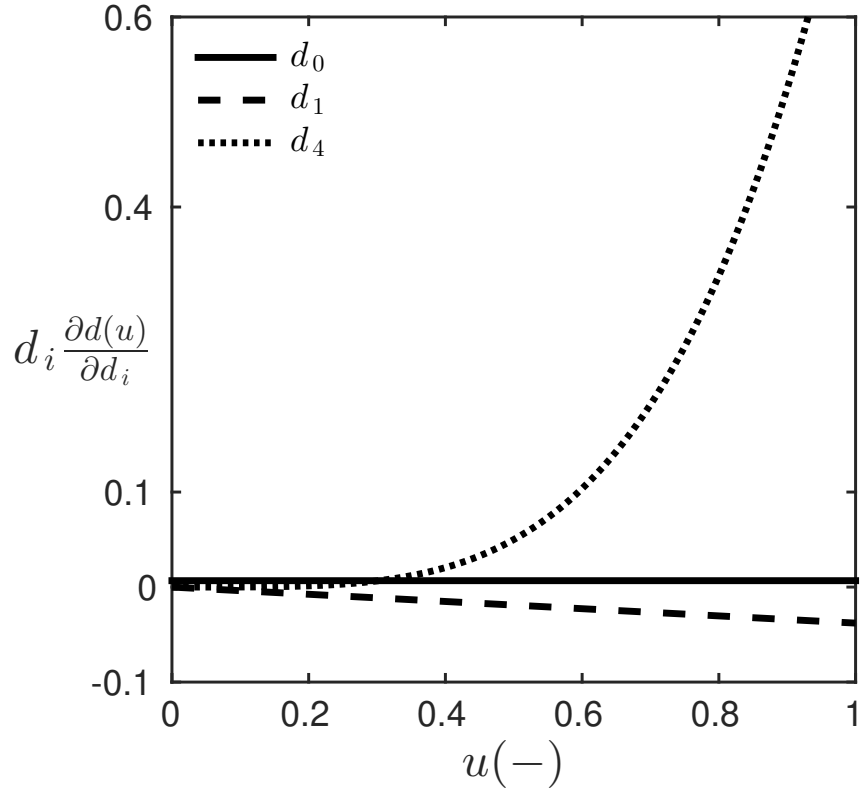}
\caption{\small\em The normalized derivative of $d\,(\,u\,)$ relative to each coefficient $d_{\,0}\,$, $d_{\,1}$ and $d_{\,4}$ }
\label{fig:derivative_D}
\end{figure}

Let us discuss which coefficient of $\,d_{\,0}\,$, $d_{\,1}\,$ or $d_{\,4}\,$ contributes the most in the context of liquid uptake or specifically $u \egal  \mathcal{O}\,(\,1\,)\,$. For this purpose, the normalized derivative  $\displaystyle{d_{\,i}\, \cdot\,\pd{d\,(\,u\,)}{d_{\,i}}}$  relative to each coefficient is taken and compared with each other. 
\begin{align*}
  d_{\,i}\, \cdot\,\pd{d\,(\,u\,)}{d_{\,i}} \egal \{\;d_{\,4}\,u^4\;;\quad d_{\,1}\,u\;;\quad d_{\,0}\;\} \,.
\end{align*}
The results are presented in Figure~\ref{fig:derivative_D}. One may conclude that coefficient $d_{\,4}$ has more influence on the value of function $d\,(\,u\,)$, when liquid uptake occurs $u \egal  \mathcal{O}(\,1\,)\,$. Therefore, in this article only coefficient $d_{\,4}$ is estimated.

\section{Comparison of the numerical predictions with experimental data}
\label{sec:parameter_estimation}

The efficiency of the proposed numerical model is demonstrated in Section~\ref{sec:numerical_validation} by comparison to reference solution. Now, the purpose is to assess the reliability of the numerical model to predict the physical phenomena. For this, the experimental observations described in previous section are used for comparison with the numerical predictions. With the given parametrization in Section~\ref{sec:parameterization_material properties}, Eq.\eqref{eq:water_uptake_dimless} writes as:
\begin{align}\label{eq:water_uptake_dimless_case_study}
  \pd{u}{t} & \egal \mathrm{Fo}\, \cdot \, \pd{}{x}\, \biggl(\,d\,(\,u\,)\, \cdot \, \pd{u}{x} \moins \mathrm{Pe} \, \cdot \,  a\,(\,u\,)\, \cdot\, u \moins \mathrm{Bo}\, \cdot \,k\,(\,u\,) \,\biggr)  \,,
\end{align} 
where
\begin{align*}
  &d\,(\,u\,)\egal d_{\,4}\,u^{\,4} \plus d_{\,1}\,u \plus d_{\,0}\,, &&a\,(\,u\,)\egal a_{\,0} \, \Bigl(\,1 \moins \int_{\,0}^{\,1} \, u\,(\,x\,,t\,) \,\mathrm{d}x \,\Bigr)\,, &&&k\,(\,u\,)\egal k_{\,3} \, u^{\,3}\,.
\end{align*}
Since literature lacks of consistent data \cite{Guizzardi2016}, the material properties, represented by functions $k\,(\,u\,)\,$, $d\,(\,u\,)$ and $a\,(\,u\,)\,$, are uncertain. Thus, parameter estimation problem are solved to determine the uncertain parameters and calibrate the numerical model with the experimental data. Within the procedure of parameter estimation, first we demonstrate the structural and practical identifiability of the three unknown parameters $\bigl(\,a_{\,0}\,,k_{\,3}\,,d_{\,4}\,\bigr)\,$. Then, the parameter estimation problem is solved and the reliability of the numerical model is discussed.


\subsection{Structural Identifiability} 
\label{seq:identif}

The aim of this section is to demonstrate the formal identifiability of the unknown parameters. The set of unknown parameters is defined as:
\begin{align*}
  \mathrm{P} \ \equiv \ \{\,\mathrm{P}_{\,i}\,\} \egal \{\,a_{\,0}\,,k_{\,3}\,,d_{\,4}\,\} \,.
\end{align*}
We assume here, $u\,(\,x\,,t\,)$ is the only observable field.

A parameter $\mathrm{P}_{\,i} \,\in\, \mathrm{P}$ is Structurally Globally Identifiable (SGI), if the following condition is satisfied \cite{Walter1982}:
\begin{align*}
  \forall t \,, \qquad u\,(\,\mathrm{P}\,) \egal u\,(\,\mathrm{P}^{\,\star}\,) \,\Rightarrow \,\mathrm{P}_{\,i} \egal \mathrm{P}^{\,\star}_{\,i} \,.
\end{align*}

In the case of our model, if $u\,(\,x\,,t\,) \ \equiv \ u^{\,\star}\,(\,x\,,t\,)$, then  $\displaystyle \pd{u}{t} \ \equiv \ \pd{u^{\,\star}}{t}$ and $\displaystyle \pd{u}{x}  \ \equiv \ \pd{u^{\,\star}}{x}$, therefore,
\begin{align}\label{eq:equality_SGI_1}
\mathrm{Fo} \, \cdot\, \pd{}{x} \, \Biggl(\, d(\,u\,)\, \cdot\,\pd{u}{x} 
\moins \mathrm{Pe}\, \cdot\,a(\,u\,)\, \cdot\,u 
\moins \mathrm{Bo}\, \cdot\,k(\,u\,)\,\Biggr) 
\equiv \\
\mathrm{Fo}\, \cdot  \, \pd{}{x} \, \Biggl(\,d^{\,\star}(\,u\,)\, \cdot\,\pd{u}{x} 
\moins \mathrm{Pe}\, \cdot\,a^{\,\star}(\,u\,)\, \cdot\,u^{\,\star} \moins \mathrm{Bo} \, \cdot\, k^{\,\star}(\,u\,)\,\Biggr) \nonumber \,.
\end{align}

In addition, straightforward replacements in the Eq.~\eqref{eq:water_uptake_dimless_case_study} shows us the following:
\begin{align*}
\pd{u}{t} \egal \mathrm{Fo} \, \cdot \,\Bigg(\, 
& \Bigl(\,4\,d_{\,4}\,u^{\,3} \plus d_{\,1}\,\Bigr)\, \cdot\pd{u}{x} 
\plus \Bigl(\,d_{\,4}\,u^{\,4} \plus d_{\,1}\,u \plus d_{\,0}\,\Bigr)
 \, \cdot\, \pd{^{\,2} u}{ x^{\,2}} \\
& \moins \mathrm{Pe}\, \cdot\,a_{\,0}\, \cdot\,
\Bigl(\,1\moins \int_{\,0}^{\,1} \, u\, \mathrm{d}x\,\Bigr)\, \cdot\,\pd{u}{x} \moins 3\,\mathrm{Bo}\, \cdot\,k_{\,3}\, \cdot\,u^{\,2} \, \cdot\,\pd{u}{x}\Bigg) \,,
\end{align*}
which shows that Equality~\eqref{eq:equality_SGI_1} becomes:
\begin{align*}
\mathrm{Fo}\, \cdot \,\Bigg(\, & \Bigl(\,4\,d_{\,4}\,u^{\,3} \plus d_{\,1} \,\Bigr) \, \cdot\, \pd{u}{x} \plus \Bigl(\, d_{\,4} \, u^{\,4} \plus d_{\,1}\,u \plus d_{\,0}\,\Bigr) \, \cdot\, \pd{^{\,2} u}{x^{\,2}} \\
& \moins \mathrm{Pe}\, \cdot\,a_{\,0}\, \cdot\,\Bigl(\,1 \moins \int_{\,0}^{\,1} \, u\, \mathrm{d}x\,\Bigr)\, \cdot\,\pd{u}{x} \moins 3\,\mathrm{Bo} \, \cdot\, k_{\,3}\, \cdot\,u^{\,2} \, \cdot\,\pd{u}{x}\Bigg) \equiv \\
& \mathrm{Fo} \, \cdot\,\Bigg(\, \Bigl(\,4\,d^{_,\star}_{\,4}\, u^{\,3} \plus d_{\,1}\,\Bigr) \, \cdot\pd{u}{x}  \plus \Bigl(\,d^{\,\star}_{\,4}\, u\,^{\,4} \plus d_{\,1}\,u \plus d_{\,0}\,\Bigr)\, \cdot\, \pd{^{\,2} u}{x^{\,2}} \\[4pt]
& \qquad \qquad \qquad \qquad \qquad \quad
\moins \mathrm{Pe}\, \cdot\,a^{\star}_{\,0}\,\Bigl(\,1\moins \int_{\,0}^{\,1} \, u\, \mathrm{d}x\,\Bigr)\, \cdot\,\pd{u}{x} 
\moins 3\,\mathrm{Bo}\, \cdot\,k^{\,\star}_{\,3}\, \cdot\,  u^{\,2}\, \cdot\,\pd{u}{x}\Bigg)\,,
\end{align*}
and the final expression is
\begin{align*}
\Bigg(\,\Bigl(\,d_{\,4}\moins d^{_,\star}_{\,4} \,\Bigr)\, \cdot\, \Bigl(\,4\,u^{\,3} \plus u^{\,4}\, \pd{^{\,2} u}{x^{\,2}} \, \Bigr) \moins \mathrm{Pe}\, \cdot\,\Bigl(\,a_{\,0}\moins a^{_,\star}_{\,0} \,\Bigr)\, \cdot\,\Bigl(\,1 \moins \int_{\,0}^{\,1} \, u\, \mathrm{d}x\,\Bigr)\, \cdot\,\pd{u}{x} \\ \moins 3\,\mathrm{Bo}\, \cdot\,\Bigl(\, k_{\,3}\moins k^{_,\star}_{\,3} \,\Bigr)\, \cdot\,u^{\,2} \, \cdot\,\pd{u}{x}\Bigg) \equiv 0
\end{align*}

Since that $\displaystyle{\Bigg\{\, \Biggl(\,4\,u^{\,3} \plus u^{\,4}\, \pd{^{\,2} u}{x^{\,2}} \, \Biggr)\,, \Biggl(\,1 \moins \int_{\,0}^{\,1} \, u\, \mathrm{d}x\,\Biggr)\, \cdot\,\pd{u}{x}\,, \,u^{\,2}\, \cdot\,\pd{u}{x}\Bigg\}}$  are independent, then $\,d_{\,4}\,$, $k_{\,3}$ and $a_{\,0}\,$ are SGI \cite{Ollivier2002}.


\subsection{Practical Identifiability}
\label{seq:pr_identif}

As demonstrated in previous Section, the three unknown parameters $\,d_{\,4}\,$, $k_{\,3}$ and $a_{\,0}\,$ are globally identifiable. This result remains theoretical. Before solving the parameter estimation problem, it is of major importance to study the sensitivity coefficients of the parameters to demonstrate the practical identifiability. If the sensitivity coefficients are either small or correlated, the estimation problem is difficult and very sensitive to measurement errors. The sensitivity coefficient is defined as the first derivative of the (numerical) observations with respect to an unknown parameter \cite{Finsterle2015, Walter1990}:
\begin{align*}
  Y_{\,P_{\,i}} \,(\,t\,) & \egal \nicefrac{\sigma_{\,p}}{\sigma_{\,H}} \; \pd{\mathrm{H}}{\mathrm{P}_{\,i}} \egal \nicefrac{\sigma_{\,p}}{\sigma_{\,H}} \; \int_{\,0}^{1}\, \,X_{\,P_{\,i}}  \mathrm{d}x \,, \quad X_{\,P_{\,i}}  \egal \pd{\mathrm{u}}{\mathrm{P}_{\,i}} \,,
\end{align*}
where $\sigma_{\,H}$ and $\sigma_{\,p}$ are scaling factors. In this case, $\sigma_{\,H}$ corresponds to the uncertainty on the experimental observations. The quantity $\sigma_{\,p}$ is set to unity since all unknown parameter have the same uncertainty on the \emph{a priori} values.

\begin{figure}
\begin{center}
\subfigure[\label{fig:sens_coefDK}]{\includegraphics[width=0.45\textwidth]{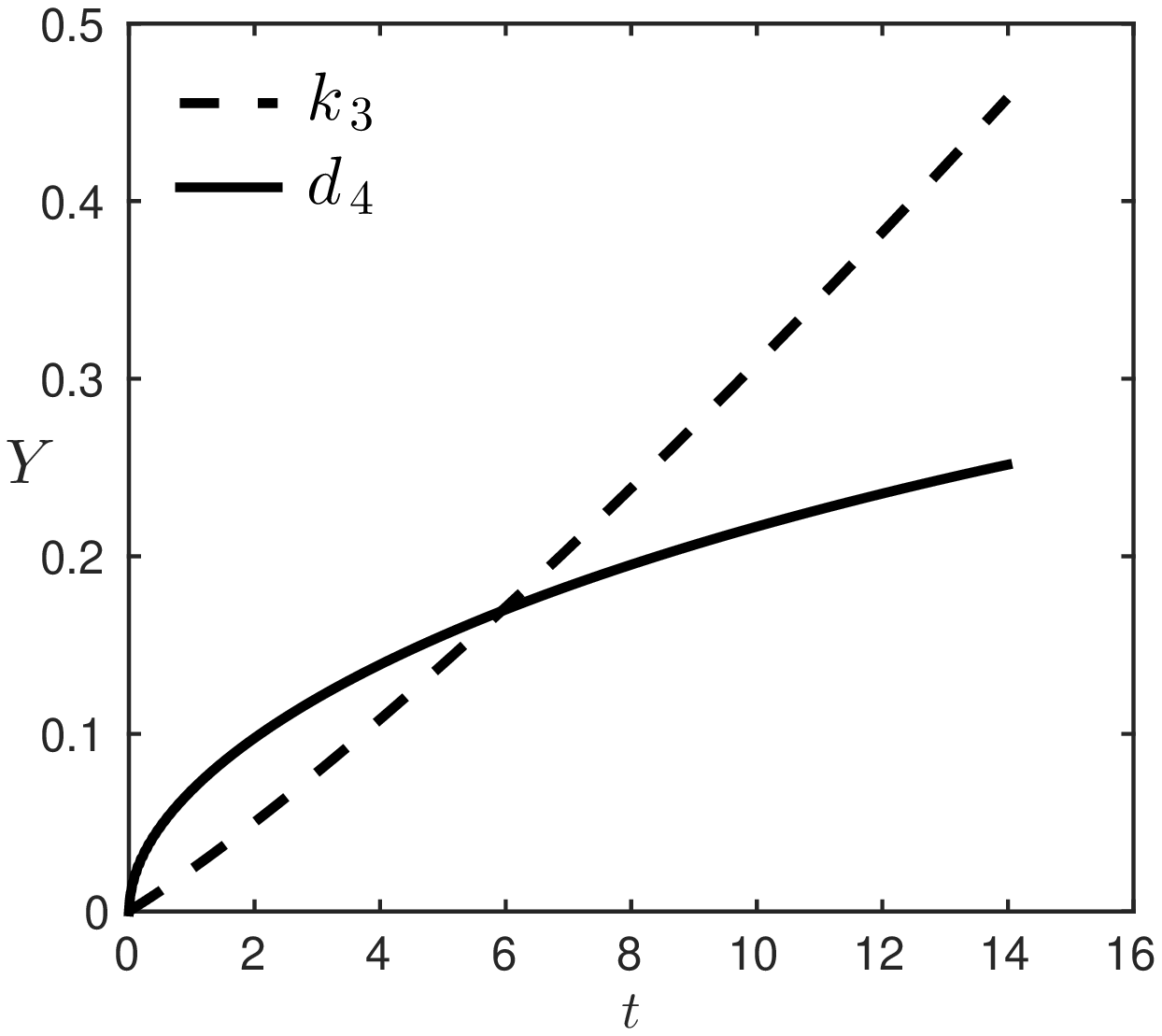}} 
\hspace{0.2cm}
\subfigure[\label{fig:sens_coefa0}]{\includegraphics[width=0.46\textwidth]{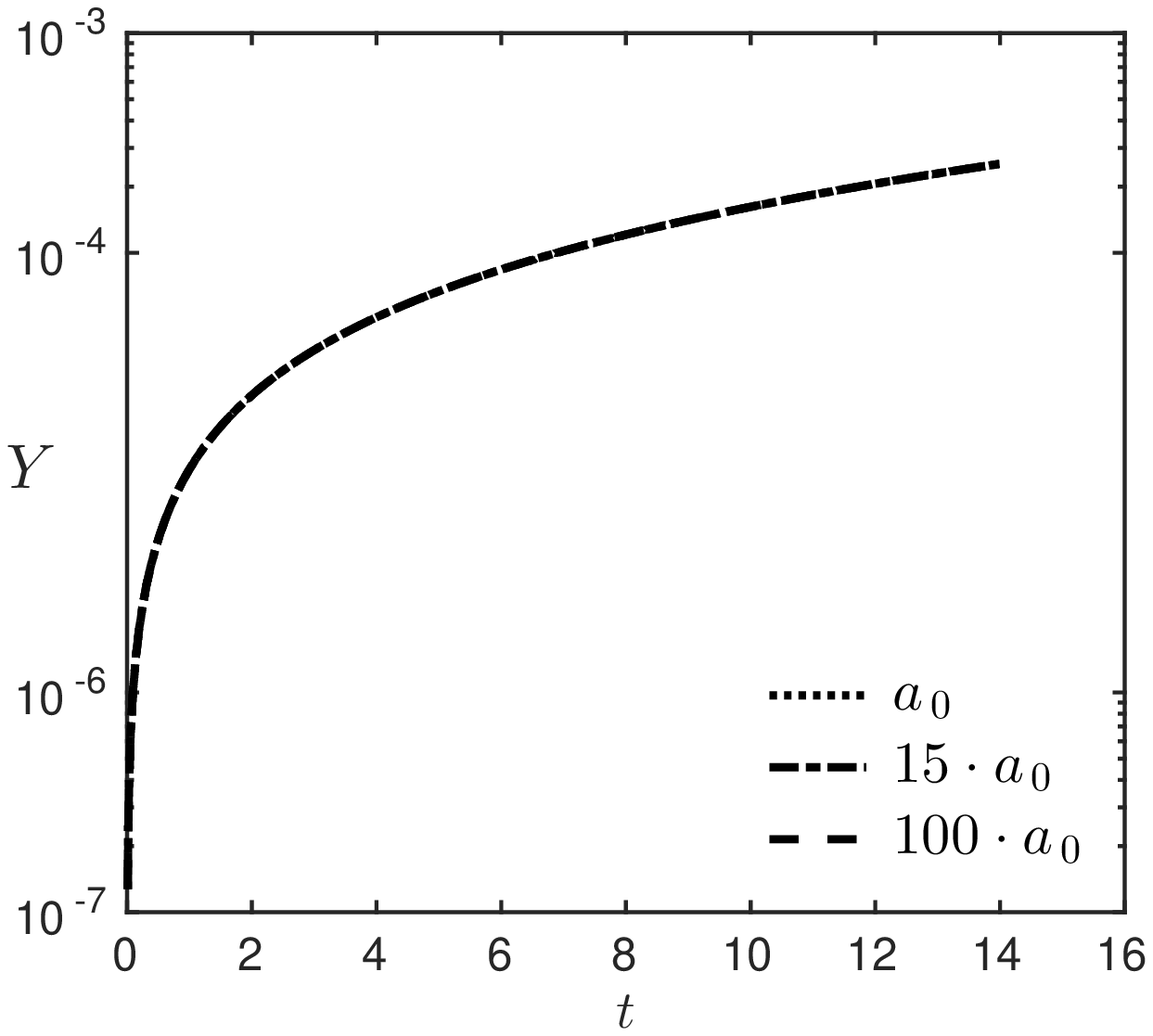}} 
\caption{\small\em Time variation of the sensitivity coefficients \emph{(a)} $Y_{\,k_{\,3}}\,$, $Y_{\,d_{\,4}}$ and \emph{(b)}  $Y_{\,a_{\,0}}\,$.}
\end{center}
\end{figure}

To compute the sensitivity coefficients, Equation~\eqref{eq:water_uptake_dimless_case_study} is differentiated with respect to each unknown parameter. Three differential equations are obtained enabling to compute the sensitivity coefficients. For the parameter $k_{\,3}\,$, we define $X_{\,k_{\,3}} \, \eqdef \, \displaystyle  \pd{u}{k_{\,3}}$ obtained by solving the following equation:  
\begin{align*}
  \pd{X_{\,k_{\,3}}}{t} \egal \mathrm{Fo}\, \cdot\,\pd{}{x}\,\Bigg( 
  & \,d\,(\,u\,)\, \cdot\,\pd{X_{\,k_{\,3}}}{x} \plus \Bigl(\,\tilde{d}\,(\,u\,)\, \cdot\,\pd{u}{x} \moins \mathrm{Pe}\, \cdot\,a\,(\,u\,) \moins \mathrm{Bo}\, \cdot\,\tilde{k}\,(\,u\,)\,\Bigr)\, \cdot\,X_{\,k_{\,3}}  \\[4pt]
  & \moins \mathrm{Bo}\,u^{\,3} \plus \mathrm{Pe}\,a_{\,0}\,\int_{\,0}^{\,1}\,X_{\,k_{\,3}}\,\mathrm{d}x\,\Bigg)\,,
\end{align*}
where
\begin{align*}
  &\tilde{d}\,(\,u\,)\egal4\,d_4\,u^3 \plus d_1\,,&&\tilde{k}\,(\,u\,)\egal3\,k_3\,u^2\,.
\end{align*}
For the parameter $d_{\,4}\,$, we define $X_{\,d_{\,4}} \, \eqdef \, \displaystyle \pd{u}{d_{\,4}}$ obtained by solving the following equation:  
\begin{align*}
  \pd{X_{\,d_{\,4}}}{t} \egal \mathrm{Fo}\, \cdot\,\pd{}{x}\,\Bigg( & \,d\,(\,u\,)\, \cdot\,\pd{X_{\,d_{\,4}}}{x} \plus \Bigl(\,\tilde{d}\,(\,u\,)\, \cdot\,\pd{u}{x} \moins \mathrm{Pe}\, \cdot\,a\,(\,u\,) \moins \mathrm{Bo}\, \cdot\,\tilde{k}\,(\,u\,)\,\Bigr)\, \cdot\,X_{\,d_{\,4}} \\[4pt]
  & \plus u^{\,4} \,\pd{u}{x} \plus \mathrm{Pe}\,a_{\,0}\,\int_{\,0}^{\,1}\,X_{\,d_{\,4}}\,\mathrm{d}x\,\Bigg)\,.
\end{align*}
And last, for the parameter $a_{\,0}\,$, we define $X_{\,a_{\,0}}  \, \eqdef \, \displaystyle \pd{u}{a_{\,0}}$ obtained by solving the following equation:  
\begin{align*}
  \pd{X_{\,a_{\,0}}}{t} \egal \mathrm{Fo}\, \cdot\,\pd{}{x}\,\Bigg( & \,d\,(\,u\,)\, \cdot\,\pd{X_{\,a_{\,0}}}{x} \plus \Bigl(\,\tilde{d}\,(\,u\,)\, \cdot\,\pd{u}{x} \moins \mathrm{Pe}\, \cdot\,a\,(\,u\,) \moins \mathrm{Bo}\, \cdot\,\tilde{k}\,(\,u\,)\,\Bigr)\, \cdot\,X_{\,a_{\,0}} \\[4pt]
  & \moins \mathrm{Pe} \, \cdot\, \biggl(\,1\moins \int_{\,0}^{\,1}\,u(\,x\,,t\,)\,\mathrm{d}x \moins \,a_{\,0}\,\int_{\,0}^{\,1}\,X_{\,a_{\,0}}\,\mathrm{d}x\,\biggr)\, \cdot\,u \,\Bigg)\,.
\end{align*}
The boundary and initial conditions are the same for each coefficient $\,X_{\,P_{\,i}}\,$:
\begin{align*}
  & X_{\,P_{\,i}}\,(\,x\egal 0\,,t\,)\egal 0\,,
  && X_{\,P_{\,i}} \,(\,x\egal 1\,,t\,)\egal 0\,,
  && X_{\,P_{\,i}}\,(\,x\,,t\egal 0\,)\egal 0\,.
\end{align*}
The three differential equations are solved with \SG ~numerical scheme. For the investigation of the practical identifiability, the \emph{a priori} parameters values are used. The time values of $Y_{\,k_{\,3}} \, \eqdef \, \displaystyle \pd{h}{k_{\,3}}$ and $Y_{\,d_{\,4}} \, \eqdef \, \displaystyle \pd{h}{d_{\,4}}\,$ are shown in Figure \ref{fig:sens_coefDK}. One may conclude that the estimation of the parameters $k_{\,3}$ and $d_{\,4}\,$ will likely have a good result since their sensitivity coefficients have large magnitudes and they are not correlated with each other. On the other hand, the time variation of the third coefficient $Y_{\,a_{\,0}}$ is given in Figure~\ref{fig:sens_coefa0}. It is not correlated with other coefficients, demonstrating that it is possible to estimate it. However, its magnitudes are very small showing that its estimation with satisfactory accuracy is a difficult task. One can argue that these observations are due to the \emph{a priori} values for the coefficient $a_{\,0}.$ However, as noticed in Figure~\ref{fig:sens_coefa0}, even multiplying by 15 or 100 times the \emph{a priori} value of $a_{\,0}\,$, the magnitude of the sensitivity coefficient almost doesn't change.

\subsection{Estimation and comparison with the experimental observations}

Since the formal and practical identifiability of the three unknown parameters $d_{\,4}\,$, $a_{\,0}$ and $k_{\,3}$ have been demonstrated, the aim is now to estimate them using the experimental observations obtained with the facility. First, the methodology to solve the inverse problem is briefly described. Then, the results are presented and the reliability of the numerical model predictions are discussed.


\subsubsection{Methodology to solve the parameter estimation problem}

The parameter estimation problem is solved by minimizing the following cost function by the least squares method:
\begin{equation*}
  \mathrm{J}\,\bigl(\,d_{\,4}\,,\,k_{\,3}\,,\,a_{\,0}\,\bigr):\egal \Bigl|\Bigl|\,H \bigl(\,d_{\,4}\,,\,k_{\,3}\,,\,a_{\,0}\,\bigr) \moins H_{\,\mathrm{exp}}\,\Bigr|\Bigr|_{\,2} \,.
\end{equation*}
The value of $H$ results from the solution of the direct problem~\eqref{eq:water_uptake_dimless_case_study} for a given set of parameters $\bigl(\,d_{\,4}\,,\,k_{\,3}\,,\,a_{\,0}\,\bigr)$. The value of $H_{\,\mathrm{exp}}$ is given by the measurements from the experimental facility and interpolated on the time grid of the numerical scheme.

The cost function $\mathrm{J}$ is minimized using function \texttt{fmincon} from the \texttt{Matlab\texttrademark} environment. This method uses the interior-point algorithm with constraints on the unknown parameters. Here, the upper and lower boundary constraints for the parameters are defined based on the preliminary calculations:
\begin{align*}
  d_{\,4} \, \in \,  \bigl[\,0,\,2\,\bigr] \,, 
  && k_{\,3} \, \in \, \bigl[\,0,\,2\,\bigr] \,, 
  && a_{\,0} \, \in \, \bigl[\,0,\,2\,\bigr] \,.
\end{align*}

To estimate the quality of the solution of parameter estimation problem, the normalized \textsc{Fisher} matrix is defined according to \cite{Karalashvili2015, Ucinski2004}:
\begin{align*}
  \mathrm{F} & \egal \Bigl[\,\mathrm{F}_{\,ij\,}\,\Bigr], \qquad \forall \,\bigl(\,i,\,j\,\bigr) \, \in \, \{ \,1,...,N_p\,\} \,, \quad 
  \mathrm{F}_{\,ij\,}  \egal \nicefrac{1}{\sigma_{\,H}^{\,2}}\, \int_{\,0}^{\,t_{\mathrm{max}}}{ Y_{\,P_{\,i}} \, Y_{\,P_{\,j}} \; \mathrm{dt}}\,,
\end{align*}
where $Y_{\,P_{\,i}}$ is the sensitivity coefficient of the solution related to the parameter $\mathrm{P}_{\,i}$, $\sigma_{\,H}$ the measurement uncertainty and $N_{\,p}$ the number of parameters. The matrix $\mathrm{F}$ measures the total sensitivity of the system for the measurements to variations of the entire set of parameters $\mathrm{P}$. Under some assumptions detailed in \cite{Walter1990}, the inverse of the \textsc{Fisher} matrix is the matrix of variance of the parameters considered as random variables of the given observable fields. In other words, it summarizes the quality of the information obtained in the parameter identification process. Thus, the inverse matrix of $\mathrm{F}$  is used to assess the estimation uncertainty by computing an error estimator for the parameter $P_{\,i}\,$:
\begin{align*}
  \eta_{\,i} \egal \sqrt{\Bigl(\,\mathrm{F}^{\,-1}\,\Bigr)_{\,i\,i}} \,.
\end{align*}
High values of $\eta_{\,i}$ indicate a possible high error during the parameter identification process.


\subsubsection{Results and discussion}

The estimation process is performed with the following discretization parameters $\Delta\,x^{\,\star}\egal 0.05\,$, $\Delta\,t^{\,\star}\egal 10^{\,-2}\,$, using the \SG ~numerical scheme and the \textsc{Adams}--\textsc{Bashforth}--\textsc{Moulton} time adaptive algorithm. The solution of the optimization problem requires the computation of the solution $u$ of Eq.~\eqref{eq:water_uptake_dimless} for each set of the parameters $(\,d_{\,4},k_{\,3},a_{\,0}\,)$.  During the optimization process for the case without pressure, the direct problem Eq.~\eqref{eq:water_uptake_dimless} is solved $114$ times. Each iteration requires approximately $100 \ \mathsf{s}$ to solve the direct problem. It results in a overall performance of $3 \ \mathsf{h}\,$. Considering the results from previous section and particularly Table~\ref{table:cpu_nonlin}, the same process using the standard finite--difference scheme and \textsc{Euler} approach would last for about $7.5 \ \mathsf{h}\,$. For the experimental data with pressure, the calculation of solution $u$ of Eq.~\eqref{eq:water_uptake_dimless} is computed almost $270$ times. The estimation takes about $7.5 \ \mathsf{h}\,$. This value is comparatively small to the $17.5 \ \mathsf{h}$ required with the model based on standard approach. The results highlights that the \SG ~numerical scheme, applying to parameter estimation problem, save important computational efforts compared to the standard approach with finite--difference scheme.

\begin{figure}
\begin{center}
\subfigure[]{\includegraphics[width=0.45\textwidth]{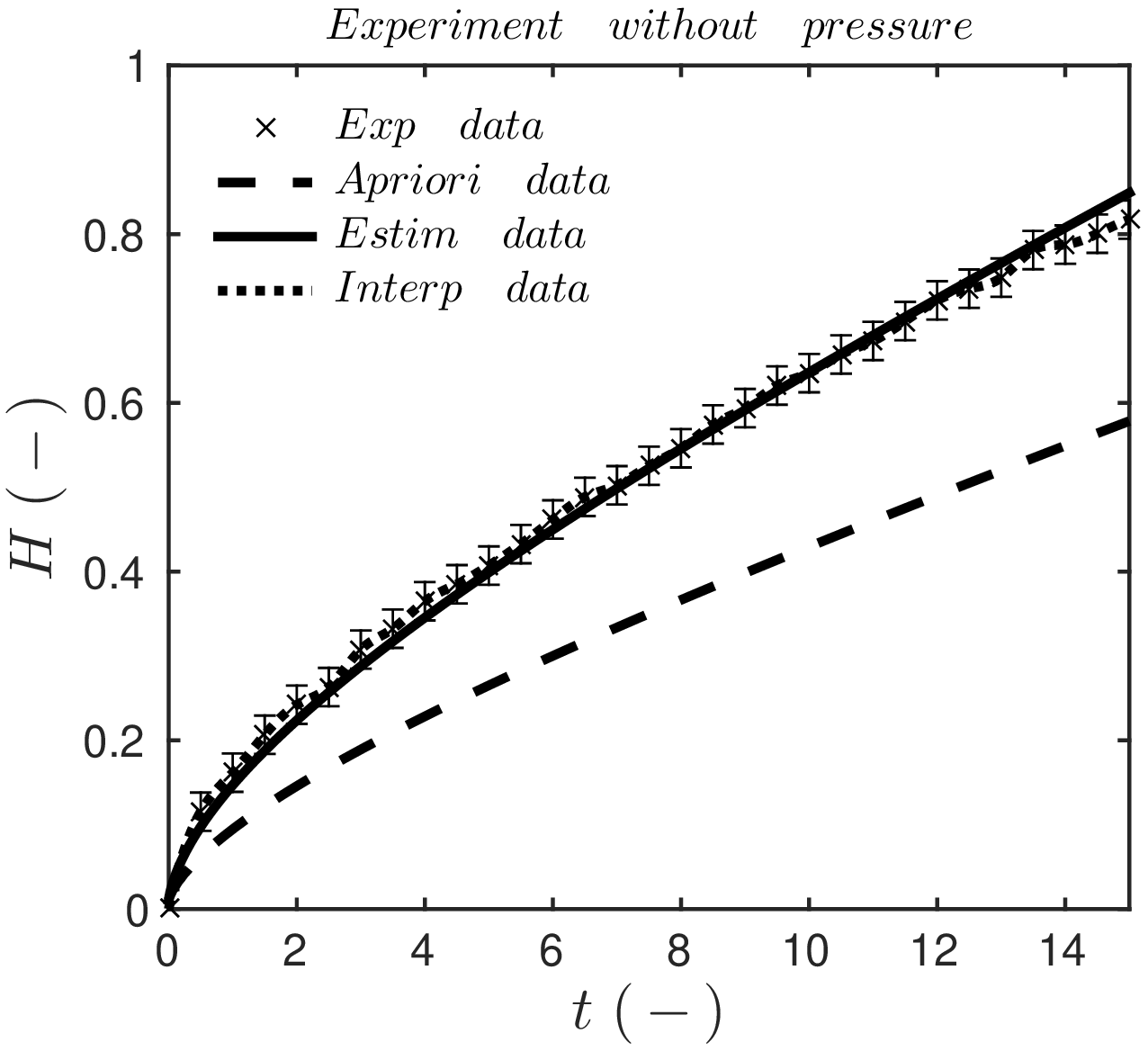}} \hspace{0.3cm}
\subfigure[]{\includegraphics[width=0.45\textwidth]{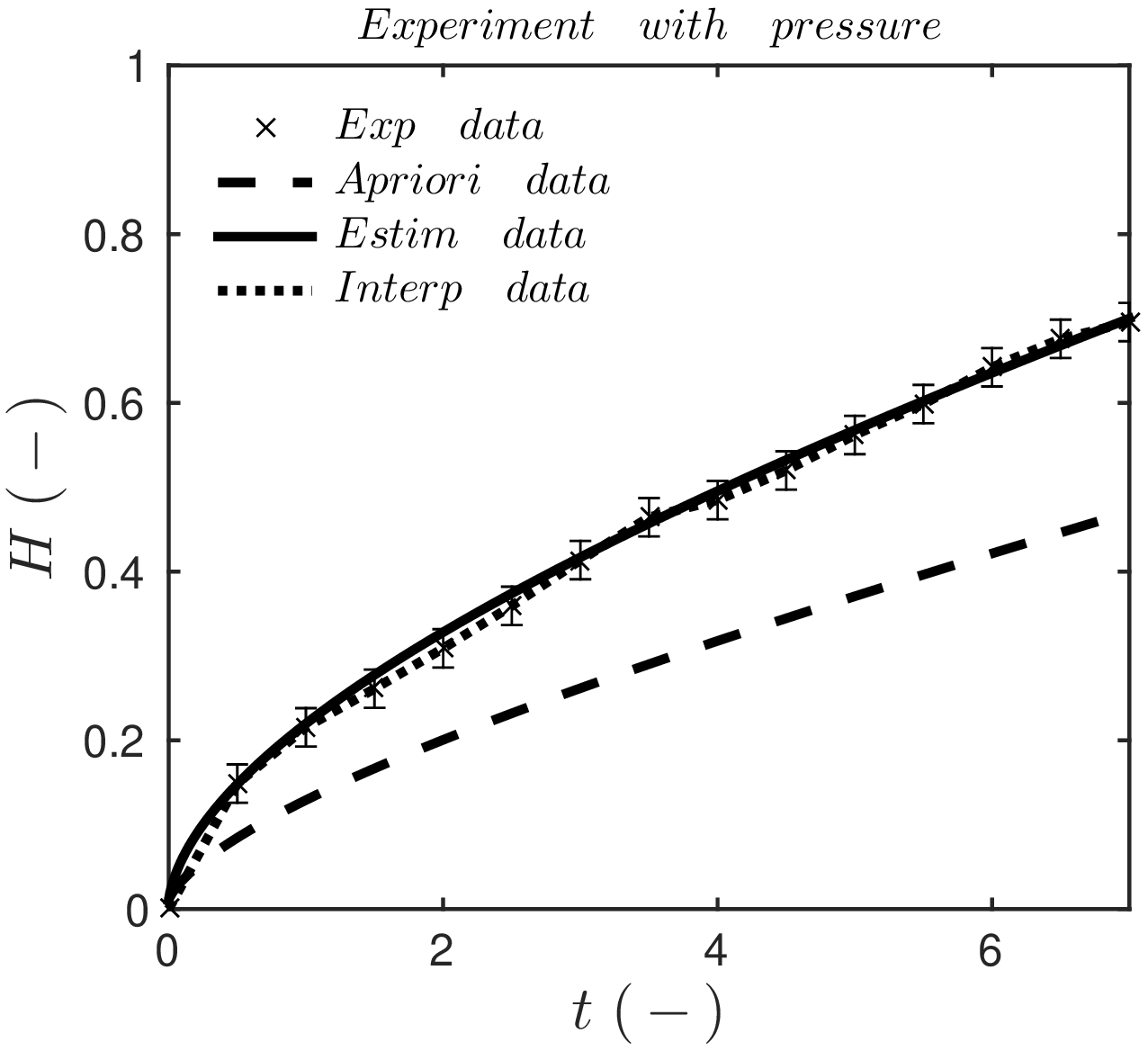}} 
\caption{\small\em Water uptake level in the brick respective to the time for case without pressure (a) and (b) case with pressure. Figure displays experimental data with uncertainty $\sigma_{\,h} \egal 0.5 \ \mathsf{cm}\,$ and its interpolation, a priori data and results with the estimated parameters with residual $\,\varepsilon_{\,2} \egal 0.05$ for (a) case and $\,\varepsilon_{\,2} \egal 0.0087$ for (b) case.}
\label{fig:height_exp}
\end{center}
\end{figure}

Figure~\ref{fig:height_exp} shows the time variation of the water height for both experiments with pressure and without pressure. It compares the solution computed using the estimated values of the parameters and the experimental data. The results demonstrate a good agreement with the experimental observations, while the solution with the \emph{a priori} values cannot approximate the experimental data well. One may say that the calibrated numerical model has a satisfactory reliability to represent the physical phenomena of liquid water uptake. Similar conclusions can be noted for the experiments with pressure.

\begin{table}
\caption{\small\em A priori and estimated values of dimensionless coefficients}
\bigskip
\begin{center}
    \begin{tabular}{ c | c | c c | c c }
    \hline
    \hline
    \multirow{2}{*}{\textbf{Parameters} }  
    & \emph{A priori}  
    & \multicolumn{2}{c|}{\textbf{No pressure}}
    & \multicolumn{2}{c}{\textbf{With pressure}} \\
    & values
    & Estimated value  
    & Error $\eta_i$ 
    & Estimated value
    & Error $\eta_i$\\ 
    \hline
    \hline
    $a_{\,0}^{\,\star}$ & 0.7 & $0.0052$ & $\pm \, 14$  & $0.55$  & $\pm \, 206$ \\ 
    $k_{\,3}^{\,\star}$ & 0.8 &$0.8257$  & $\pm \, 0.01$ &$1.3$  & $\pm \, 0.21$ \\
    $d_{\,4}^{\,\star}$ & 0.6 &$1.0$  & $\pm \, 0.002$ &$1.72$  & $\pm \, 0.02$  \\
    \hline
    Residual $\,\varepsilon_{\,2}$ & 1 &0.05 & &0.0087&
    \\
    \hline
    \end{tabular}
    \bigskip
\label{tab:param_table}
\end{center}
\end{table}

As a result of the optimization process, the estimated values, the residuals of the cost function and the error estimators are reported in the Table~\ref{tab:param_table}. The error estimator is small for parameters $k_{\,3}\,$ and $d_{\,4}\,$, proving that the accuracy of estimation is decent. However, the high values of the error estimator of parameter  $a_{\,0}\,$ indicate an unsatisfactory estimation with a very high uncertainty. These results are consistent with ones obtained when analyzing the sensitivity coefficient with \emph{a priori} values in Section~\ref{seq:pr_identif}. In addition, as noticed in Table~\ref{tab:param_table}, there is no significant difference between the \emph{a priori} and estimated values for parameters. It validated the investigations on practical identifiability carried in Section~\ref{seq:pr_identif}, since it is eligible in the local area of \emph{a priori} values.

\begin{figure}
\centering
\includegraphics[width=0.75\linewidth]{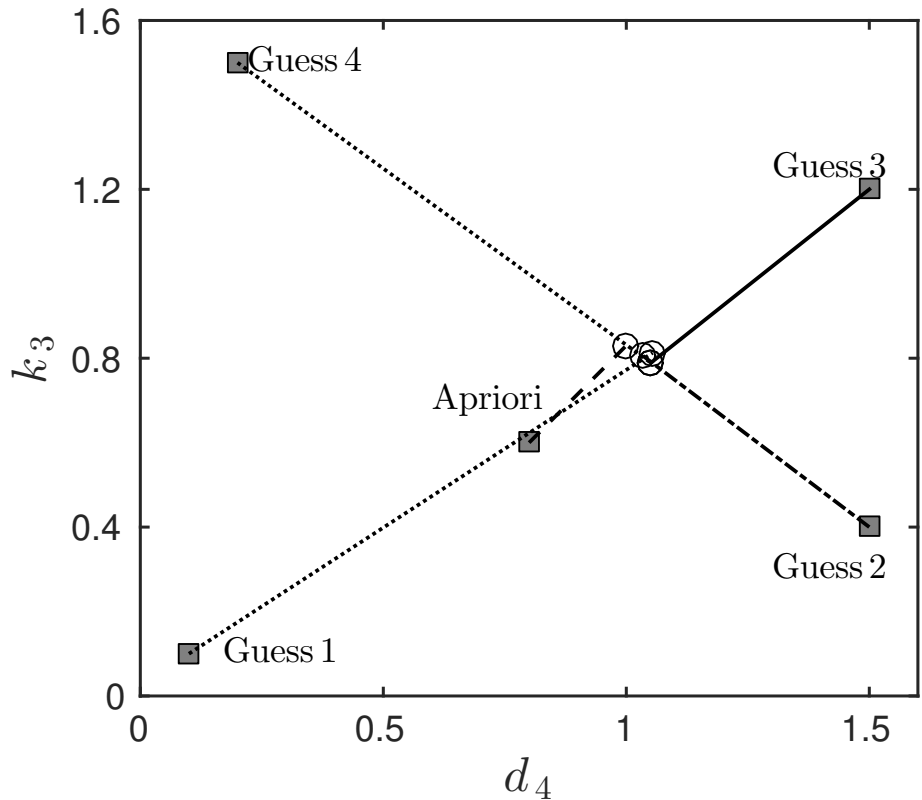}
\caption{\small\em Illustration of optimization procedure convergence.}
\label{fig:est_history}
\end{figure}

In order to demonstrate the convergence of the optimization algorithm, the following procedure is implemented. The parameter $a_{\,0}\,$ cannot be estimated accurately so its value is fixed for this analysis. Then several values of initial guesses for parameters $d_{\,4}\,$ and $k_{\,3}\,$ are taken. The estimation procedure is performed for each case. Figure~\ref{fig:est_history} shows the path from each starting point until the estimated parameters values. As one may conclude the final points are in a similar range for all the tested initial guesses. It validates the convergence of the optimization method. It should be noted that the figure does not show the intermediary computations of the algorithm procedure.

\begin{figure}
\begin{center}
\subfigure[\label{fig:theta_NPx}]{\includegraphics[width=0.45\textwidth]{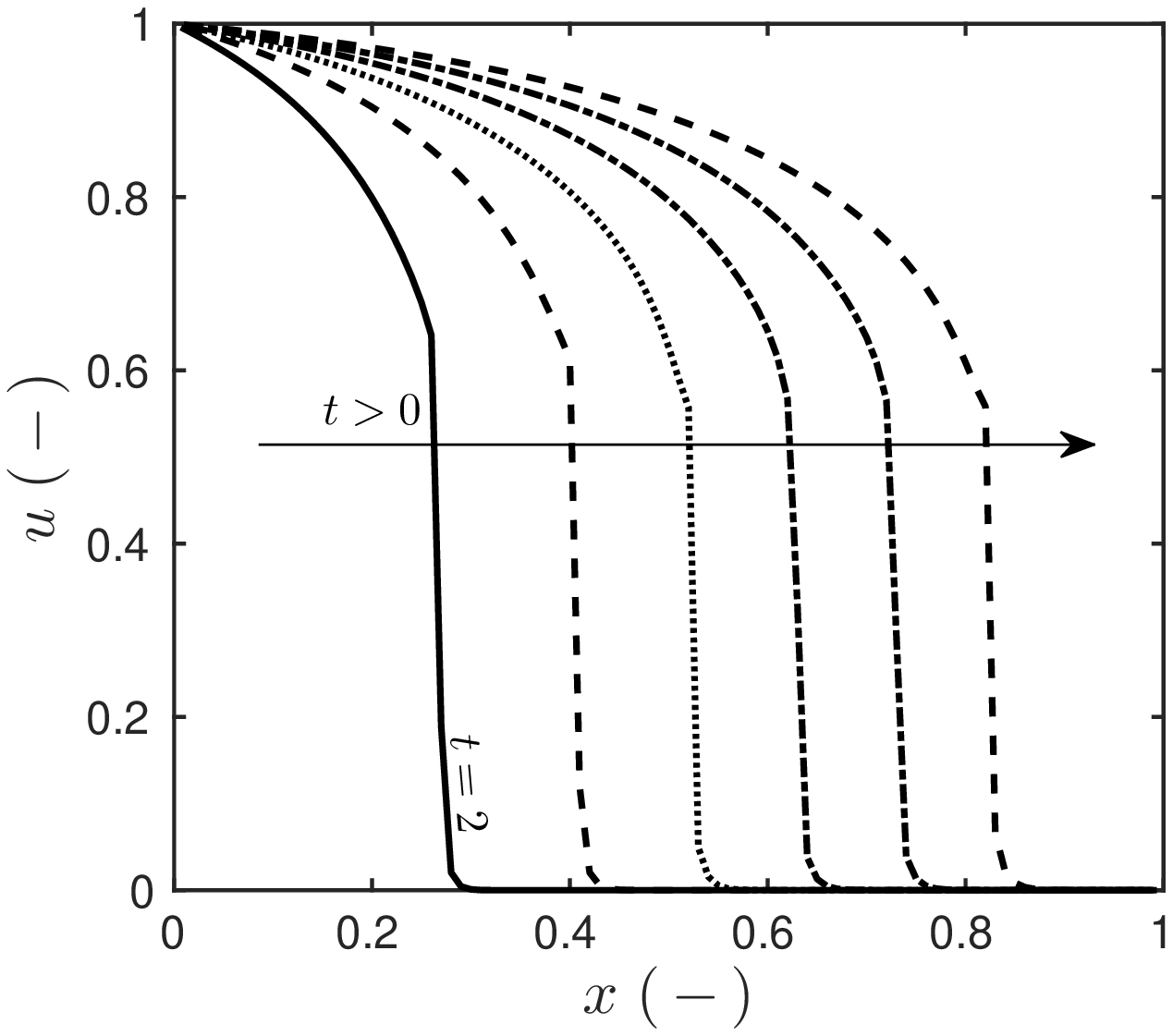}}
\hspace{0.2cm}
\subfigure[\label{fig:theta_NPt}]{\includegraphics[width=0.45\textwidth]{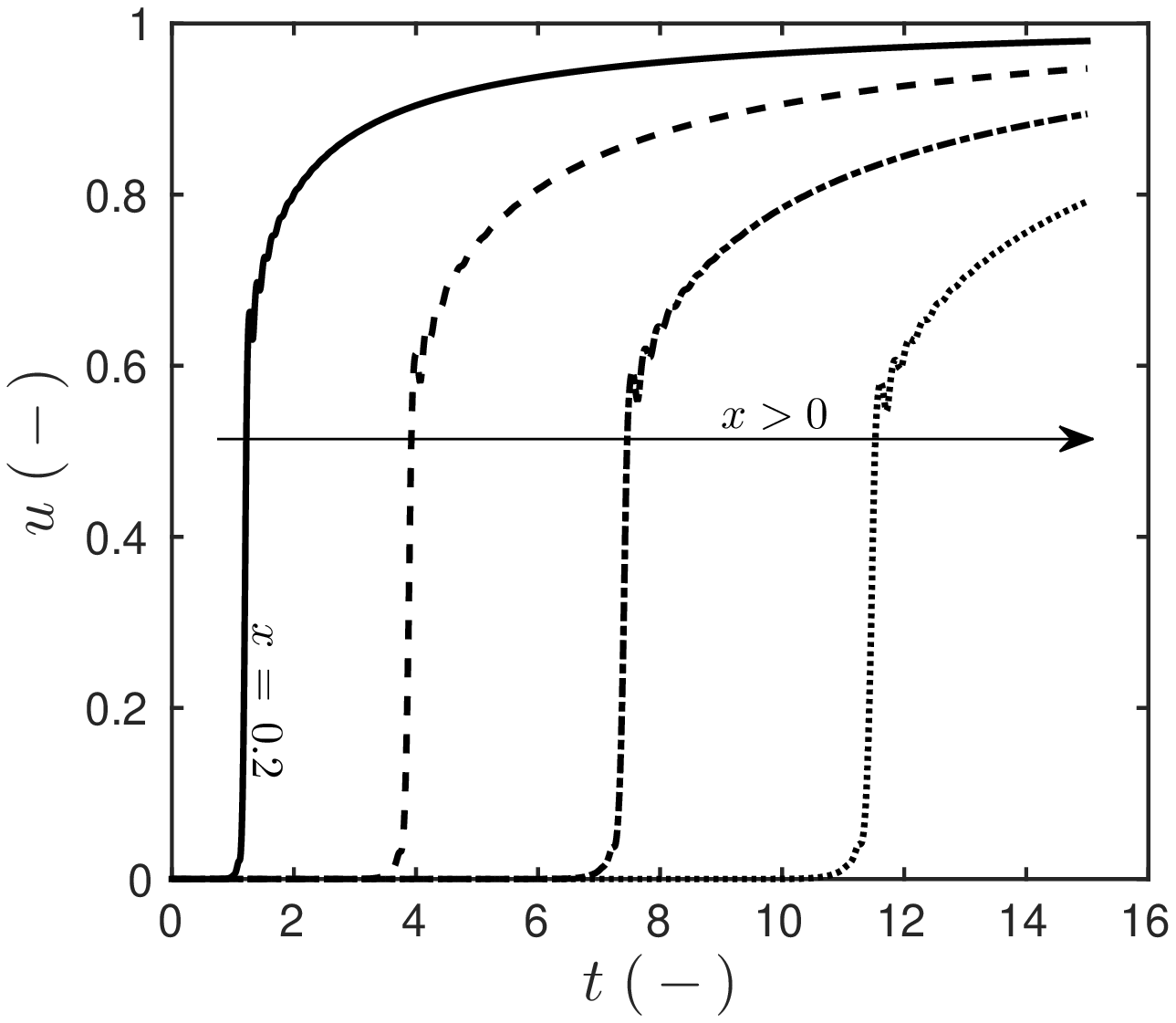}} 
\subfigure[\label{fig:theta_Px}]{\includegraphics[width=0.45\textwidth]{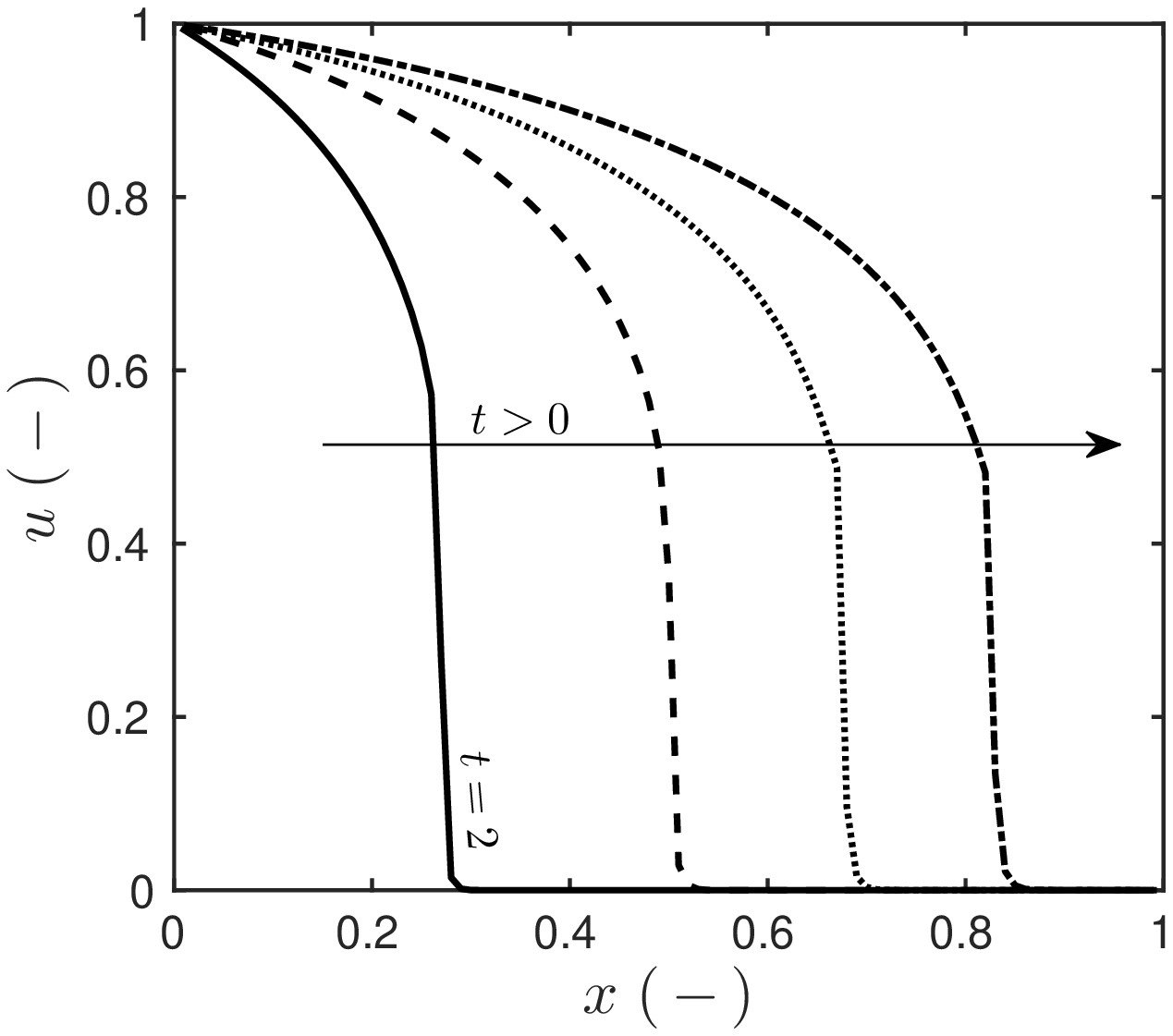}} \hspace{0.3cm}
\subfigure[\label{fig:theta_Pt}]{\includegraphics[width=0.45\textwidth]{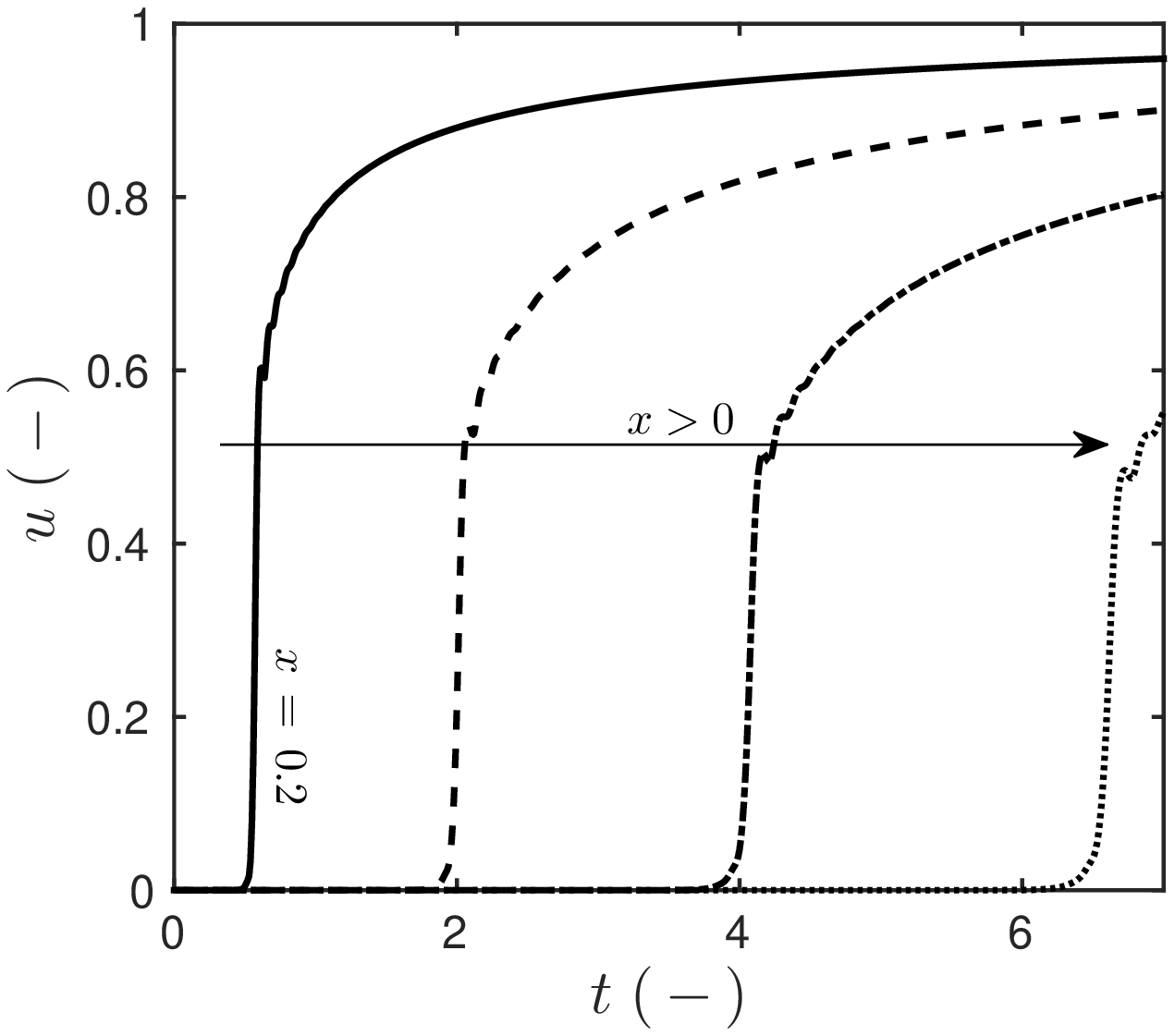}} 
\caption{\small\em Variation of the dimensionless amount of liquid water $u$ in case with and without pressure. Figures (a) and (b) displays case without pressure, relative to the height with $t \,\in\, \{\,2\,,4\,,6\,,8\,,10\,,12\,\}\,$, and  time with $x \,\in\, \{\,0.2\,,0.4\,,0.6\,,0.8\,\}\,$ respectively. Figures (c) and (d) displays case with pressure, relative to the height with time $t \,\in\, \{\,2\,,4\,,6\,,7\,\}\,$, and  time with $x \,\in\, \{\,0.2\,,0.4\,,0.6\,,0.8\,\}\,$ respectively. }
\label{fig:theta_P}
\end{center}
\end{figure}

Figure~\ref{fig:theta_P} displays the volume of liquid water relative to space and time when the brick is exposed to the open air or to a pressure difference of $-\,50 \ \mathsf{Pa}\,$. The propagation of the water front through the brick seems faster in the case of the pressure difference. As noted in Figure~\ref{fig:height_exp}, the experimental data with pressure show that the velocity of water uptake is faster since the same height of the water is achieved twice faster than with the experiment without pressure. One may conclude that the pressure influences the velocity of water uptake. However, as reported in Table ~\ref{tab:param_table}, results from the optimization problem gives different values for the set of parameters $(\,d_{\,4} \,, k_{\,3} \,,a_{\,0}\,)$ with and without pressure. Particularly, values for parameters $\,d_{\,4}\,$ and $\,k_{\,3}\,$ vary from one case to another. It may indicate that material properties may be different between the two bricks used during the experiments. Moreover, the error estimators of the advection coefficients are really poor indicating that for the given value of pressure difference, it has a minor influence on the water uptake. It should also be noted that the sensitivity coefficients of parameter $a_{\,0}$ has really low magnitudes of variations as shown in Figure~\ref{fig:sens_coefa0}.

\begin{figure}
\centering
\includegraphics[width=0.7\linewidth]{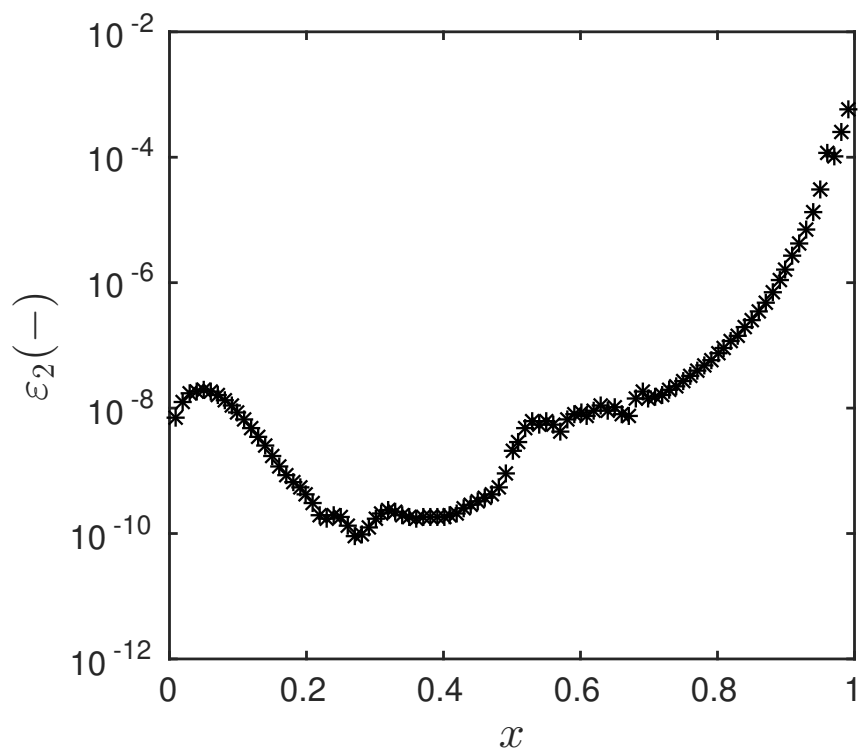}
\caption{\small\em Error $\varepsilon_{\,2}$ between solutions with \textsc{Dirichlet} and homogeneous \textsc{Neumann} boundary condition at $x \egal 1 \,$.}
\label{fig:eps2_BC}
\end{figure}

In Figure~\ref{fig:height_exp}, it can be noticed that the full length of the brick is not reached, validating the chosen physical model with \textsc{Dirichlet} boundary conditions presented in Section~\ref{seq:phys_model}. One may argue that other types of boundary conditions could have been chosen at the top of the brick $x \egal 1\,$. The \textsc{Robin}-type boundary condition is avoided. Indeed the boundary flow is proportional to the ambient and surface conditions:
\begin{align*}
   \biggl(\,d\,(\,u\,)\, \cdot \, \pd{u}{x} \moins \Pe\, \cdot \,  a_{\,0}\, \cdot\,\Bigl(\,1 \moins \int_{\,0}^{\,1} \, u\,(\,x\,,t\,)\, \mathrm{d}x\,\Bigr)\, \cdot\,u &\moins \Bo\, \cdot \,k\,(\,u\,) \,\biggr)\\
   \egal & \mathrm{Bi} \, \cdot\, \Bigl(\, u \moins u_{\,\infty} \,\Bigr)\,, \qquad x \egal 1 \,,
\end{align*}
where $\mathrm{Bi}$ is a dimensionless surface transfer coefficient and $u_{\,\infty}$ are the ambient conditions in the facility. It implies to estimate this coefficient $\mathrm{Bi}$ which is rather difficult experimentally. It can be estimated by adding the coefficient $\mathrm{Bi}$ in the inverse problem, at the price of additional computational costs. In addition, the expected values of the surface coefficient are slow and the boundary conditions can be modified into a homogeneous \textsc{Neumann} one:
\begin{align}\label{eq:BC_neumann}
  \biggl(\,d\,(\,u\,)\, \cdot \, \pd{u}{x} \moins \Pe \, \cdot\,  a_{\,0}\, \cdot\,\Bigl(\,1 \moins \int_{\,0}^{\,1} \, u\,(\,x\,,t\,)\, \mathrm{d}x\,\Bigr)\, \cdot\,u \moins \Bo\, \cdot \,k\,(\,u\,) \,\biggr) \egal & 0  \,, \nonumber \\ 
  x \egal & 1 \,,
\end{align}
To investigate the importance of modifying the definition of the boundary conditions at $x \egal 1\,$, additional computation are performed considering the problem Eq.~\eqref{eq:water_uptake_dimless_case_study} with the estimated parameters $(\,d_{\,4} \,, k_{\,3} \,,a_{\,0}\,) $ and the boundary condition Eq.~\eqref{eq:BC_neumann}. The boundary condition at $x \egal 0$ and the initial condition are unchanged:
\begin{align*}
  u(\,x\egal0\,,\,t) \ &= \ \begin{cases} 0 \,, &t\egal 0 \\1\,, &t\ > \ 0  \end{cases}\,, 
  && u(\,x\,,\,t\egal 0)  \ = \ 0 \,.
\end{align*}
Figure~\ref{fig:eps2_BC} displays the difference between the solutions computed with \textsc{Dirichlet} and \textsc{Neumann} boundary conditions. The error $\varepsilon_{\,2}$ is lower than $10^{\,-4}$ and one may conclude that the boundary conditions do not change significantly the solution of the numerical model. These results are consistent when analyzing Figures~\ref{fig:theta_Px} and \ref{fig:theta_NPx}. It can be observed that the flow at $x \egal 1$ remains null.

\begin{figure}
\centering
\includegraphics[width=0.7\textwidth]{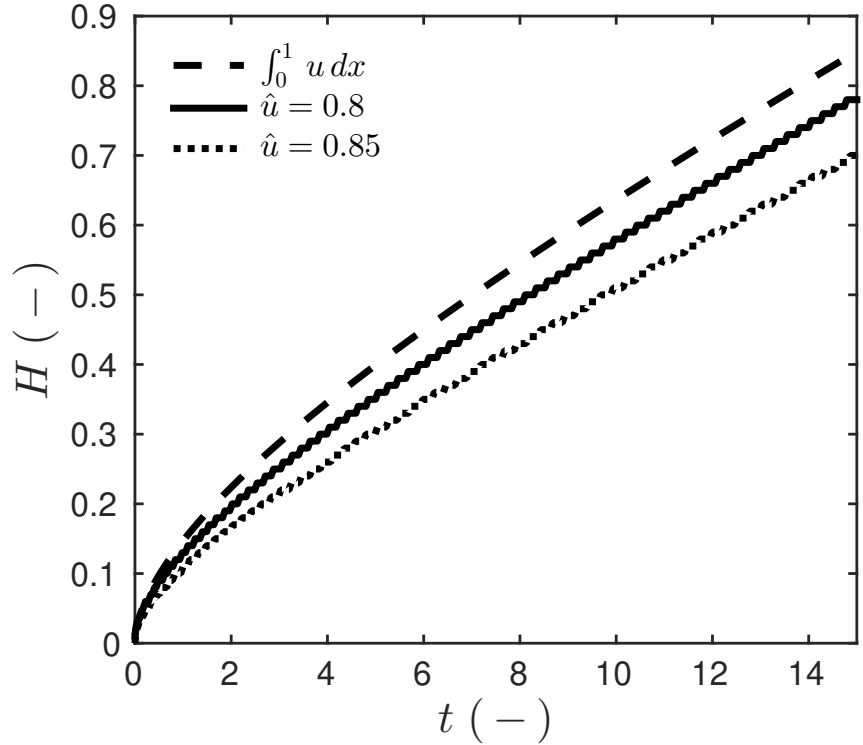}
\caption{\small\em Influence of the definition of the water front height of the brick on the numerical predictions.}
\label{fig:definition_h}
\end{figure}

Another open question is the definition of the liquid height. Previously, the height is defined by the integral of $u$ over the space domain. An alternative definition is to the height of the water front in the brick using a threshold:
\begin{align}\label{eq:definition_H2}
  H \egal \max \,\Bigl\{\, \hat{x}: \, u\,(\,\hat{x}\,,t\,) \ \geq \ \hat{u} \,\Bigr\}\,,
\end{align}
where $\hat{u}$ is a chosen value of the amount of liquid. Figure~\ref{fig:definition_h} compares the numerical solution according to the definition of $H$ with Eq.~\eqref{eq:definition_H} or \eqref{eq:definition_H2}. The numerical predictions are varying with the definition. On one hand, the integral definition is a more general approach but it assumes that the water vapor is negligible in the material. On the other hand, the definition~\eqref{eq:definition_H2} requires to determine the value of the threshold. For these reasons, the definition of the liquid height in the brick according remains an open question.

\section{Conclusion}

The moisture in the walls greatly affects the overall building performance. Such a process as rising damp has a strong impact on the energy consumption of the building. In this work, the physical model represents the water uptake process in a single brick, based on diffusion, advection and gravity fluxes. To investigate these physical phenomena it is important to have efficient numerical models in terms of computational cost and accuracy. These models require also to have a good reliability to represent the physical phenomena. To answer this issue, an innovative approach is proposed based on the \SG ~numerical scheme. The first part of the paper studies the properties of the numerical model such as accuracy,  stability conditions and computational time with a reference solution for both linear and nonlinear cases in Section~\ref{sec:numerical_validation}. The investigations show that the \SG ~numerical model is more accurate and faster than the standard approach based on finite--differences and \Eu ~explicit scheme. The \SG ~numerical model has an explicit formulation avoiding costly sub-iterations at each time step. Since its stability condition is nonlinear and scales with $\Delta t \ \simeq \ \Delta x$ for large space discretization, the \SG ~numerical scheme combined with an adaptive time step approach is particularly efficient to save computational efforts.

Along with numerical methods, the reliability of the proposed numerical model is analyzed by comparing the numerical predictions to experimental observations. The experimental facility is presented in Section~\ref{sec:exp_facility}. Empirical data are gathered through a set of two experiments. First, the height of the rising front is obtained for a brick under normal conditions. Then, the height front is measured in another brick submitted to a pressure of $-\,50 \ \mathsf{Pa}\,$. Then, the comparison between numerical simulations and the experimental observations of the height of the rising front is carried out in Section~\ref{sec:parameter_estimation}. Since the literature lacks consistent data for the material properties \cite{Guizzardi2016}, the uncertain parameters are determined by solving a parameter estimation problems. Before solving the  parameter estimation problem, the structural and practical identifiability of the three unknown parameters are demonstrated in Sections~\ref{seq:identif} and \ref{seq:pr_identif}, respectively. The results of the parameter estimation problem show that the diffusivity and liquid conductivity are estimated with satisfactory accuracy. The advection coefficient cannot be estimated with accuracy even with a pressure difference of $-\,50 \ \mathsf{Pa}$ at the top of the brick. The sensitivity of the numerical solution with respect to this parameter is much smaller compared to the diffusivity and liquid conductivity. With the estimated parameters, there is a satisfactory agreement between the numerical predictions and the experimental observations, highlighting a good reliability of the model. Important computational efforts are saved thanks to the efficiency of the numerical model. The \SG ~numerical scheme enables to save by $50\%$ the computational cost compared to the standard approach.

Even if this work enhances the efficiency of the \SG ~numerical scheme for the solution of advection--diffusion equation with gravity flux, the rising damp problem require simulation of the physical phenomena in $2-$dimensions. Thus, further research should be conducted to extend the numerical model in this way. 


\section*{Acknowledgements}

This work was partly funded by the ``Conseil \textsc{Savoie Mont Blanc}'' (CSMB), the French Atomic and Alternative Energy Center (CEA) and the French Environment and Energy Management Agency (ADEME) (through the research program CAPVENT). The authors also acknowledge the Junior Chair Research program ``Building performance assessment, evaluation and enhancement'' from the University of \textsc{Savoie Mont Blanc} in collaboration with the French Atomic and Alternative Energy Center (CEA) and Scientific and Technical Center for Buildings (CSTB). The authors also would like to acknowledge Dr. B.~\textsc{Rysbaiuly} for his valuable discussions.


\appendix
\section{Demonstration of uniqueness of the solution}
\label{sec:uniqueness_solution}

This section proposes to demonstrate the uniqueness of the solution of liquid uptake in porous media. Since $\; u \,(\,x\,,t\,) \, > \, 0 \;$ at $\;  t \, > \, 0 \,, $ equation~\eqref{eq:main_eq} can be transformed as
\begin{align*}
  \pd {u}{t} \ &= \ \pd{}{x}\, \Biggl(\, d\;\pd{u}{x} \moins u\, \Bigl(\,a \plus \nicefrac{k(\,u\,)}{u}\,\Bigr) \,\Biggr)\,,
\end{align*}
then regularized into a general advection--diffusion equation:
\begin{align}\label{eq:adv_diff}
  \pd {u}{t} \ &= \ \pd{}{x}\, \Biggl(\, d\,\pd{u}{x} \moins \tilde{a} \,u \,\Biggr)\,,
  && x \, \in \, \Omega \egal \bigl[\,0 \,,\, 1 \, \bigr] \,, \; t \, > \, 0 \,,
\end{align}
where $\tilde{a}\,(\,u\,) \, \eqdef \, \biggl(\, a \plus \frac{k\,(\,u\,)}{u} \biggr)\,$. Further in the Section, we assume the linearized representation $\tilde{a}$ of $\tilde{a}\,(\,u\,)\,$ around $u_{\,0}\,$:
\begin{align*}
  \tilde{a} \egal a \plus \dfrac{k(\,u_{\,0}\,)}{u_{\,0}}\,,  
\end{align*} 
where $u_{\,0}$ is a given function of $\,(\,x\,,t\,)\,$.

The following boundary and initial conditions are considered:
\begin{align}\label{eq:adv_diff_CL_IC}
  u\,(\,0\,,\,t\,) \egal \uinf^{\,L}\,(\,t\,)  \,, 
  && u\,(\,1\,,\,t\,) \egal  \uinf^{\,R}\,(\,t\,)  \,, 
  && u\,(\,x\,,\,0\,) \egal 0 \,.
\end{align}
In order to study the uniqueness of the solution, we use the so-called \emph{energy method} \cite{Evans2010}. For that purpose, we assume that two solutions $u_{\,1}\,(\,x\,,\,t\,)$ and $u_{\,2}\,(\,x\,,\,t\,)$ satisfy Eqs.~\eqref{eq:adv_diff} and \eqref{eq:adv_diff_CL_IC}. We define $w(\,x\,,\,t\,)$ as:
\begin{align*}
  w\,(\,x\,,\,t\,) \, \eqdef \, u_{\,1}\,(\,x\,,\,t\,) \moins u_{\,2}\,(\,x\,,\,t\,) \,,
\end{align*}
solution of the equation:
\begin{align}\label{eq:adv_diff_w}
  \pd{w}{t} \egal \pd{}{x} \, \biggl(\, d \, \pd{w}{x} \moins \tilde{a} \, w \,\biggr) \,,
\end{align}
with the boundary and initial conditions:
\begin{align}\label{eq:adv_diff_CL_w}
  w\,(\,1\,,\,t\,) \egal 0 \,, 
  && w\,(\,0\,,\,t\,) \egal 0 \,,
  && w\,(\,x\,,\,0\,) \egal 0 \,.
\end{align}

If the energy $E\,(\,t\,) \, \eqdef \, \displaystyle \int_{\Omega} \, w^{\,2} \,(\,x\,,\,t\,) \, \mathrm{d}x$ is decreasing and $E\,(\,0\,) \egal 0\,$, then the solution of Eqs~\eqref{eq:adv_diff} and \eqref{eq:adv_diff_CL_IC} is unique :
\begin{align*}
  \od{E}{t} \, \leqslant \, 0 \,.
\end{align*}
To study the validity of this condition, both sides of Eq.~\eqref{eq:adv_diff_w} are multiplied by $w$ and integrated over $\Omega\,$:
\begin{align*}
\int_{\Omega} \,w \, \pd{w}{t} \, \mathrm{d}x
\egal \int_{\Omega} \, w \,  \pd{}{x} \, \biggl(\, d \, \pd{w}{x} \moins \tilde{a} \, w \,\biggr) \, \mathrm{d}x \,,
\end{align*}
which by differentiating under the integral becomes:
\begin{align*}
\frac{1}{2} \, \od{E}{t} \egal \int_{\Omega} \, w \,  \pd{}{x} \, \biggl(\, d \, \pd{w}{x} \moins \tilde{a} \, w \,\biggr) \, \mathrm{d}x \,,
\end{align*}
By performing an integration by parts, we obtain:
\begin{align*}
\frac{1}{2} \, \od{E}{t} \egal
\Biggl[\, w \, \biggl(\, d \, \pd{w}{x} \moins \tilde{a} \, w \,\biggr) \,\Biggr]_{\,x \egal 0}^{\,x \egal 1} 
\moins \int_{\Omega} \, \biggl(\, d \, \pd{w}{x} \moins \tilde{a} \, w \,\biggr) \, \pd{w}{x} \, \mathrm{d}x \,,
\end{align*}
which becomes using the boundary conditions~\eqref{eq:adv_diff_CL_w}:
\begin{align}
\label{eq:inequality_1}
\frac{1}{2} \, \od{E}{t} \egal
\moins \int_{\Omega} \, d \, \biggl(\,\pd{w}{x}\,\biggr)^{\,2} 
\plus \int_{\Omega} \, \tilde{a} \, w \, \pd{w}{x} \, \mathrm{d}x \,.
\end{align}
The \textsc{Cauchy}--\textsc{Schwartz} inequality writes:
\begin{align*}
\Biggl|\, \int_{\Omega} \, f \, g \, \Biggr| \, \leqslant \, 
\Biggl(\, \int_{\Omega} \, f^{\,2} \, \mathrm{d}x \,\Biggr)^{\,\half} \, 
\Biggl(\, \int_{\Omega} \, g^{\,2} \, \mathrm{d}x \,\Biggr)^{\,\half} \,,
\end{align*}
for $f$ and $g$ square--integrable realx--value functions. Thus, the inequality~\eqref{eq:inequality_1} yields to:
\begin{align}\label{eq:inequality_2}
  \, \frac{1}{2} \, \od{E}{t} \,  \, \leqslant \, \moins \int_{\Omega} \, d \, \biggl(\,\pd{w}{x}\,\biggr)^{\,2} \plus \tilde{a} \, \Biggl(\, \int_{\Omega} \, w^{\,2} \mathrm{d}x \,\Biggr)^{\,\half} \, \Biggl(\, \int_{\Omega} \, \biggl(\, \pd{w}{x}\,\biggr)^{\,2}  \mathrm{d}x \,\Biggr)^{\,\half} \,.
\end{align}
For $u \, \in \, W_{\,0}^{\,1\,,\,p}\,$, the \textsc{Poincar\'e} inequality states:
\begin{align*}
  \Bigl|\Bigl|\, u \,\Bigr|\Bigr|_{\,L^{\,p}\,(\,\Omega\,)} \, \leqslant \, C_{_{\,\Omega}} \; \Bigl|\Bigl|\, \grad \, u \,\Bigr|\Bigr|_{\,L^{\,p}\,(\,\Omega\,)} \,,
\end{align*}
where $C_{_{\,\Omega}}$ is a constant depending on $p$ and $\Omega\,$ only.  Thus, for a one-dimensional problem and for $p \egal 2\,$, the inequality~\eqref{eq:inequality_2} becomes:
\begin{align*}
  \, \frac{1}{2} \, \od{E}{t} \,  \, \leqslant \, \moins \int_{\Omega} \, d \, \biggl(\,\pd{w}{x}\,\biggr)^{\,2} \plus \tilde{a} \, C_{_{\,\Omega}} \, \int_{\Omega} \, \biggl(\, \pd{w}{x}\,\biggr)^{\,2}  \mathrm{d}x \,,
\end{align*}
which can be re-written as:
\begin{align*}
  \, \frac{1}{2} \, \od{E}{t} \,  \, \leqslant \, \moins \Bigl(\, d \moins \tilde{a} \, C_{_{\,\Omega}} \,\Bigr) \, \int_{\Omega}  \, \biggl(\,\pd{w}{x}\,\biggr)^{\,2} \,.
\end{align*}
Therefore, we have uniqueness of the solution of Eqs~\eqref{eq:adv_diff} and \eqref{eq:adv_diff_CL_IC} if $k \moins \tilde{a} \, C_{_{\,\Omega}} \, \geqslant \, 0 \,$. In other terms, the generalized advection coefficient has to be small compared to the diffusion one to have unique solution.


\bigskip\bigskip
\addcontentsline{toc}{section}{References}
\bibliographystyle{abbrv}
\bibliography{biblio}
\bigskip\bigskip

\end{document}